\documentclass[fleqn,usenatbib]{mnras}

\usepackage{newtxtext,newtxmath}
\usepackage[T1]{fontenc}
\usepackage{ae,aecompl}
\usepackage{graphicx}	
\usepackage{xspace}
\usepackage[table]{xcolor}

\newcommand{\vpeak}{\ensuremath{V_{\rm peak}}\xspace}
\newcommand{\vmax}{\ensuremath{V_{\rm max}}\xspace}
\newcommand{\Mpeak}{\ensuremath{M_{\rm peak}}\xspace}

\newcommand{\Mstar}{\ensuremath{M_*}\xspace}
\newcommand{\fstar}{\ensuremath{f_*(t_f = 1\,{\rm Gyr})}\xspace}

\newcommand{\Msun}{\ensuremath{M_\odot}\xspace}

\definecolor{hpurple}{HTML}{7E16DF}

\newcommand{\logMpeak}{\ensuremath{\log_{10} (M_{\rm peak}[h^{-1}\mathrm{M}_\odot])}\xspace}
\newcommand{\logMstar}{\ensuremath{\log_{10} (M_*[\mathrm{M}_\odot])}\xspace}

\newcommand{\galaxpy}{\textsc{galaxpy}\xspace}
\newcommand{\galaxev}{\textsc{galaxev}\xspace}
\newcommand{\umachine}{\textsc{universemachine}\xspace}
\newcommand{\illustris}{IllustrisTNG\xspace}

\definecolor{Gray}{gray}{0.9}
\newcolumntype{a}{>{\columncolor{Gray}}c}

\newcommand{\bit}{\begin{enumerate}}
\newcommand{\eit}{\end{enumerate}}

\title[The colour of star formation]{Surrogate modelling the Baryonic Universe I: The colour of star formation}
\author[Chaves-Montero \& Hearin]{
Jon\'{a}s Chaves-Montero$^{1}$\thanks{E-mail: \href{mailto:jchavesmontero@anl.gov}{jchavesmontero@anl.gov}}
and Andrew Hearin$^{1}$
\\
$^{1}$ HEP Division, Argonne National Laboratory, 9700 South Cass Avenue, Lemont, IL 60439, USA.
}

\date{Accepted XXX. Received YYY; in original form ZZZ}
\pubyear{2019}

\begin{document}
\label{firstpage}
\pagerange{\pageref{firstpage}--\pageref{lastpage}}
\maketitle

\begin{abstract}
The spectral energy distribution of a galaxy emerges from the complex interplay of many physical ingredients, including its star formation history (SFH), metallicity evolution, and dust properties. Using {\sc galaxpy}, a new galaxy spectral prediction tool, and SFHs predicted by the empirical model {\sc universemachine} and the cosmological hydrodynamical simulation IllustrisTNG, we isolate the influence of SFH on optical and near-infrared colours from 3\,200 to 10\,800 {\AA} at $z=0$. By carrying out a principal component analysis, we show that physically-motivated SFH variations modify galaxy colours along a single direction in colour space: the SFH-direction. We find that the projection of a galaxy's present-day colours onto the SFH-direction is almost completely regulated by the fraction of stellar mass that the galaxy formed over the last billion years. Together with cosmic downsizing, this results in galaxies becoming redder as their host halo mass increases. We additionally study the change in galaxy colours due to variations in metallicity, dust attenuation, and nebular emission lines, finding that these properties vary broad-band colours along distinct directions in colour space relative to the SFH-direction. Finally, we show that the colours of low-redshift SDSS galaxies span an ellipsoid with significant extent along two independent dimensions, and that the SFH-direction is well-aligned with the major axis of this ellipsoid. Our analysis supports the conclusion that variations in star formation history are the dominant influence on present-day galaxy colours, and that the nature of this influence is strikingly simple.
\end{abstract}



\section{Introduction}
\label{sec:intro}

The spectral energy distribution (SED) of a galaxy encodes detailed information about the physics of galaxy formation, as star formation history \citep{heavens_etal00}, metallicity evolution \citep{tojeiro_etal07}, dust properties \citep{conroy_etal10a}, initial mass function \citep{Conroy2009}, nebular emission \citep{wilkins_etal13}, as well as other galaxy properties, act in concert to shape observed SEDs. Due to these correlations, inferring the physical properties of individual galaxies is a long-standing challenge in astronomy \citep{kauffmann_etal03, acquaviva_etal11, Conroy2013, chevallard_charlot16}. In this work, we study how the physical ingredients of galaxy formation manifest in photometric observations of galaxies, with a special focus on the influence of star formation history (SFH) on broad-band optical and near-infrared galaxy colours.

The relationship between SFH and colours has been studied previously in the literature using parametric approaches, \citep{iyer_gawiser17, Carnall2019}, non-parametric approaches \citep{leja_etal19, iyer_etal19}, simulations \citep{Nelson2018a}, and semi-analytic models \citep{Pacifici2012}. Here, we leverage the \umachine galaxy evolution model \citep{Behroozi2019} and the \illustris cosmological hydrodynamical simulation \citep{Pillepich2018a} as generators of SFHs that are well-motivated both physically and observationally. Our aim is to identify the characteristic signature(s) that SFH imprints upon the statistical distribution of galaxy colours.

We start by studying the properties of SFHs drawn from \umachine and \illustris. Even though SFHs predicted by these models present different shapes and levels of burstiness, we find that these SFHs share a very important property: galaxies in more massive haloes reach the peak of their SFH earlier, present a larger stellar mass, and quench faster. This well-known effect is commonly known as {\em downsizing} \citep[e.g.,][]{Cowie1996}. There is plenty of observational evidence for downsizing \citep[e.g.,][and references therein]{Fontanot2009}, which is a general prediction of structure formation theories, hydrodynamical simulations \citep[e.g.,][]{Dave2016}, and abundance matching \citep{conroy_wechsler09}.

After getting rid of unwanted correlations between the SFH and other galaxy features, we use \galaxpy, a new open-source python-based galaxy spectral prediction tool, and SFHs drawn from \umachine and \illustris to produce broad-band colours that isolate the influence of physically-motivated SFH variations. We then carry out a Principal Component Analysis (PCA) of these colours to study how physically motivated SFH variations, as well as changes in other galaxy properties, manifest in observations of broad-band optical and near-infrared colours. PCA has been widely used in the literature in closely related contexts, such as spectral classification of galaxies \citep{Connolly1995} and quasars \citep{Yip2004}, stellar mass estimation \citep{Chen2012}, K-corrections \citep{Blanton2007}, and inferring spectra from optical imaging \citep{kalmbach_connolly17}. Finally, we seek to determine the galaxy properties with the strongest influence on galaxy colours by carrying out a PCA on the colours of a volume-limited sample of galaxies from the Sloan Digital Sky Survey \citep[SDSS,][]{york00, dawson13}. Our work complements the results shown in \citet{Pacifici2015}, who studied the impact of various galaxy properties on colours.

This work is the first in a series building a new approach to simulation-based forward-modelling of the galaxy-halo connection: Surrogate modelling the Baryonic Universe (SBU). Our present scope is somewhat pedagogical; we demarcate the specific influence of star formation history on galaxy colours to improve our understanding of the degeneracies between model ingredients. One of the main conclusions of this work is that the colours of a galaxy primarily depend upon the fraction of stellar mass formed over the last billion years. Taking advantage of this important simplification, in a companion paper to the present work we build a computationally efficient surrogate model for the prediction of broad-band galaxy colours. These two papers set up the basis of SBU, a model that we will use to produce synthetic skies by populating cosmological simulations with galaxies. Forward-modelling galaxy catalogues is essential for ongoing experiments such as the Dark Energy Survey\footnote{\url{https://www.darkenergysurvey.org/}} \citep[DES,][]{des}, the Hyper-Suprime Cam\footnote{\url{https://hsc.mtk.nao.ac.jp/ssp/}} \citep[HSC,][]{hsc}, and the Kilo-Degree Survey\footnote{\url{http://kids.strw.leidenuniv.nl/}} \citep[KiDS,][]{kids}, and near-future surveys like the Javalambre Physics of the Accelerating Universe Astrophysical Survey\footnote{\url{http://www.j-pas.org/}} \citep[J-PAS,][]{jpas}, the Large Synoptic Survey Telescope\footnote{\url{http://www.lsst.org/}} \citep[LSST,][]{ivezic_etal08, lsst_science_book}, the Spectro-Photometer for the History of the Universe, Epoch of Reionization and Ices Explorer\footnote{\url{http://spherex.caltech.edu/}} \citep[SPHEREX,][]{spherex}, and the Wide Field Infrared Survey Telescope \citep[WFIRST,][]{wfirst}. Our aim is to apply the SBU approach to mimic the measurements of these and other surveys, thereby enabling a precise characterisation of the impact of systematic uncertainties on cosmological parameters, as well as setting constraints on galaxy formation physics using large-scale structure data.

The paper is organised as follows. In \S\ref{sec:data} we first describe the main features of the newly developed model \galaxpy, and then we briefly present the most important characteristics of \umachine and \illustris. Using SFHs drawn from these two models, in \S\ref{sec:SFHhalo} we study the connection between galaxies and dark matter haloes. Then, in \S\ref{sec:SFHSED} we study the influence of SFH on broad-band SEDs, and in \S\ref{sec:colorsgprop} we address the impact of initial mass function, metallicity, dust attenuation, and nebular emission on galaxy colours. We discuss our findings in the context of the existing literature in \S\ref{sec:discussion}, and we conclude with a summary of our principal results in \S\ref{sec:conclusions}.

Throughout this paper we use {\it Planck} 2\,015 cosmological parameters \citep{planck14b}: $\Omega_{\rm m}= 0.314$, $\Omega_\Lambda = 0.686$, $\Omega_{\rm b} = 0.049$, $\sigma_8 = 0.83$, $h = 0.67$, and $n_{\rm s} = 0.96$. Magnitudes of observed and simulated galaxies are in the AB system, while halo and stellar masses are in $h^{-1}M_{\odot}$ and $M_{\odot}$ units, respectively. Simulated colours are computed by convolving \galaxpy spectra with LSST transmission curves\footnote{\url{https://github.com/lsst/throughputs/tree/master/baseline/}}. For the sake of brevity, some figures in the main body of the paper display results for \umachine data only; for completeness, we include analogous results for \illustris in Appendix~\S\ref{app:tng_res}.


\section{Datasets and models}
\label{sec:data}

In this section, we first detail the main features of \galaxpy, a new publicly available galaxy spectral prediction tool. Then, we briefly introduce the data-driven model \umachine and the cosmological hydrodynamical simulation \illustris, from which we draw physically-motivated star formation histories.


\subsection{GALAXPY}
\label{sub:data_galaxpy}

The spectral energy distribution of a galaxy depends on many ingredients, including its star formation history, metallicity, dust properties, and nebular emission. Due to complex correlations between these, modelling galaxy SEDs is a long-standing problem in astronomy. The most followed approach to model SEDs is Stellar Population Synthesis (SPS), which relies on stellar evolution theory to produce precise SEDs typically in the UV, optical, and near-infrared range \citep[for a review see][]{Conroy2013}. The basic unit of SPS models is a Simple Stellar Population (SSP), i.e. a set of stars with the same metallicity created by an instantaneous star-forming event; the motivation behind this approach is that any stellar population can be written as a sum of SSPs. We can find many examples of SPS models in the literature such as \galaxev \citep{Bruzual2003}, \citet{Maraston2005}, and \textsc{fsps} \citep{Conroy2009}. In this section, we introduce a new open-source python-based galaxy spectral prediction tool, \galaxpy\footnote{\url{https://github.com/jchavesmontero/galaxpy/}}, which uses \galaxev for constructing SSPs, accounts for attenuation by dust following the prescriptions outlined in \citet{Charlot2000}, and models nebular emission lines as \citet{Gutkin2016}. We proceed to briefly introduce the main properties of \galaxpy.

The most common technique used by SPS codes to produce SEDs for SSPs is isochrone synthesis \citep{Charlot1991, Bruzual1993}, which relies on three main ingredients: stellar evolutionary tracks, spectral libraries, and initial mass functions (IMFs). The first accounts for the evolution of stars with the same mass, age, and metallicity; in \galaxpy we use Padova 1994 evolutionary tracks describing stars with metallicities $Z=0.0001$, 0.0004, 0.004, 0.008, 0.02 ($Z_\odot$), and 0.05 \citep{Alongi1993, Bressan1993, Fagotto1994a, Fagotto1994b, Fagotto1994c, Girardi1996}. The second ingredient is responsible for producing SEDs using the output of stellar evolutionary tracks; our model relies on the widely used BaSeL 3.1 library \citep{Westera2002}. Finally, the IMF controls the distribution of stellar masses in a population of newborn stars. We will consider \citet{Salpeter1955}, \citet{Kroupa2001}, and \citet{Chabrier2003} IMFs.

To account for the attenuation of starlight by dust, in \galaxpy we follow the prescriptions outlined in \citet{Charlot2000}. This dust model attenuates the light coming from SSPs by a factor $\exp[\tau_\lambda(t)]$ given by

\begin{equation}
    \label{eq:dust}
    \tau_\lambda(t) = 
    \begin{cases} 
        \tau_{V,{\rm y}}(\lambda/\lambda_0)^n & \text{for } t \leq 10\,\mathrm{Myr},\\
        \tau_{V,{\rm o}}(\lambda/\lambda_0)^n & \text{for } t > 10\,\mathrm{Myr},
    \end{cases}
\end{equation}

\noindent where $\tau_{V,{\rm y}}$ ($\tau_{V,{\rm o}}$) controls the attenuation of light coming from stars younger (older) than 10 Myr, $\lambda_0=5500\,${\AA}, and $n=-0.7$. This approach is inspired by observations: optical depths measured from emission lines, which depend mostly on light emitted by young stars, are approximately two times larger than those determined using the underlying continuum \citep{Calzetti1994}, which is largely controlled by old stars. Note that this model neglects the absorption of ionizing photons by dust in star forming regions. Regarding attenuation by the intergalactic medium, wavelengths shortward of Lyman-$\alpha$ are attenuated according to the recipes introduced in \citet{Fan2006} and \citet{Becker2015}.

Nebular emission lines are modelled following the same approach as in \citet{Gutkin2016}, which assumes that the nebular emission of an entire galaxy is proportional to its star formation rate. This model includes six free parameters that control emission line ratios: interstellar metallicity $Z_{\rm ISM}$, zero-age ionisation at the Str\"{o}mgren radius $U_S$, dust-to-metal mass ratio $\xi_{\rm d}$, carbon-to-oxygen abundance ratio $C/O$, hydrogen gas density $n_{\rm H}$, and high-mass cutoff of the IMF $m_{\rm up}$. For simplicity, throughout this work we only consider variations in $Z_{\rm ISM}$ and $U_S$, as these are the two parameters that most strongly affect emission line ratios. The first regulates gas cooling \citep[e.g.,][]{Spitzer1978}, which for $Z_{\rm ISM}>0.006$ ($Z_{\rm ISM}<0.006$) is dominated by collisionally excited (fine-structure) transitions. The second controls the size of H{\sc II} regions, which become more compact as $U_S$ grows, thereby increasing the ratio of high- to low-ionisation lines. We set the other parameters to the best-fitting values determined in \citet{Gutkin2016} using low redshift spectroscopic data from the SDSS survey: $\xi_{\rm d}=0.3$, $C/O=(C/O)_\odot=0.44$, $n_{\rm H}=100\,\mathrm{cm}^{-3}$, and $m_{\rm up}=100\,\Msun$. Once emission lines are produced, we also attenuate these according to Eq.~\ref{eq:dust}. Note that nebular continuum emission is currently neglected in \galaxpy.

Throughout most of this work we use the same \galaxpy configuration: Chabrier IMF, $Z=0.02$ (solar metallicity), $\tau_{V,{\rm y}}=1$, $\tau_{V,{\rm o}}=0.3$, and no nebular emission lines. If emission lines are considered, the fiducial values of the two free emission line parameters are $Z_{\rm ISM}=0.001$ and $\log_{10}U_S=-3.5$.


\subsection{SFH library}
\label{sub:data_sfh_library}

One of our most important aims is to study the influence of SFH variations on broad-band galaxy colours. In principle, the star formation history of a galaxy can be inferred from observations; however, even from high-resolution spectra most algorithms can only recover smooth histories following simple functional forms, or alternatively deliver results at a reduced number of control points (see \S\ref{sec:discussion}). Consequently, SFHs estimated from observations either present a reduced temporal resolution or may be biased. To avoid these issues, we resort to physically motivated SFHs from simulations. The main drawback of this approach is that our predictions become model dependent; to mitigate the dependence of prescriptions of a certain galaxy formation model on the results, we will study SFHs from both the empirical model \umachine \citep{Behroozi2019} and the cosmological hydrodynamical simulation \illustris \citep{Marinacci2018, Naiman2018, Nelson2018b, Pillepich2018a, Springel2018}. Note that these two models follow completely distinct approaches to simulate galaxies. We detail the main properties of \umachine and \illustris below.


\subsubsection{UniverseMachine}
\label{sub:data_UMachine}

\umachine is a data-driven model of galaxy evolution that maps SFHs onto dark matter (sub)halos empirically \citep{Behroozi2019}. Assuming that each (sub)halo contains a single galaxy, this model assigns to each galaxy a star formation rate (SFR) that depends on the maximum circular velocity ever attained by its (sub)halo, \vpeak, and its redshift. Using Conditional Abundance Matching \citep{hearin_etal14}, \umachine allows for a possible correlation between SFR and the relative change in the maximum circular velocity of a (sub)halo over the last dynamical time, $\Delta\vert_{\tau_{\rm dyn}}\vmax$, where the strength of this correlation is parameterized as a function of both halo mass and redshift. The stellar mass of each galaxy is determined by integrating the SFH of its main progenitor, while additionally accounting for mass gained through mergers or lost via passive evolution. As shown in \citet{Behroozi2019}, the best-fitting \umachine model captures a wide range of statistics summarising the observed galaxy distribution, e.g. stellar mass functions, quenched fractions, and two-point galaxy clustering.

We generate the \umachine data used in the present work by running the publicly available code\footnote{\url{https://bitbucket.org/pbehroozi/universemachine/}} on merger trees identified in the Bolshoi-Planck simulation with Rockstar and ConsistentTrees \citep{behroozi_etal13a, behroozi_etal13b, klypin_etal16, rodriguez_puebla_etal16}. The Bolshoi-Planck simulation was carried out using the ART code \citep{kravtsov_etal97}. This simulation evolved $2048^3$ dark-matter particles of mass $m_{\rm p}=1.35\times10^{8}M_{\odot}$ on a simulation box of $250\ h^{-1}{\rm Mpc}$ on a side under cosmological parameters closely matching \citet{planck14b}. After running \umachine, we end up with $\simeq700\,000$ galaxies with SFHs tabulated at each of the 178 publicly available Bolshoi-Planck snapshots. 

It has been long-known that the distribution of galaxy colours presents a strong bimodality \citep[e.g.,][]{Strateva2001, Baldry2004}, where galaxies with blue (red) colours typically present high (low) star formation rates. Many approaches have been proposed to classify galaxies into these two populations \citep[e.g.,][]{Pozzetti2010}; we will consider galaxies with specific SFRs above and below $\log_{10}({\rm SFR}/\Mstar[{\rm yr}^{-1}])=-11$ as star-forming and quenched, respectively. In this work, we will study star-forming and quenched galaxies separately.

Given that the evolution of satellite galaxies inside haloes is highly model dependent, we will restrict our analysis to central galaxies. This election assumes that the difference between central and satellite SEDs is primarily driven by differences in quenching times, and so we have adopted this simplifying assumption to reduce the dependence of our findings on the particularities of each galaxy formation model. We avoid selecting backsplash satellites -- centrals that were satellites in the past -- by imposing that all present-day centrals cannot have any progenitor classified as a satellite. We find that approximately 400\,000 \umachine galaxies satisfy these criteria. From this sample, we select the 1\,000 star-forming and quenched galaxies with host halo masses closest to $\logMpeak=11.5$, 12, 12.5, and 13, respectively, where \Mpeak is the maximum mass that a (sub)halo has ever attained.


\subsubsection{IllustrisTNG}

\illustris is a suite of cosmological hydrodynamical simulations carried out using the moving-mesh code {\sc Arepo} \citep{Springel2010}. Incorporating a comprehensive galaxy formation model with radiative gas cooling, star formation, galactic winds, and AGN feedback \citep{Weinberger2017, Pillepich2018a}, \illustris solves for the joint evolution of dark matter, gas, stars, and supermassive black holes from $z=127$ up to present time. We will use publicly available data from the largest hydrodynamical simulation of the suite, TNG300-1 \citep{Nelson2018a}, which evolved $2\,500^3$ gas tracers together with the same number of dark matter particles in a periodic box of $302.6$ Mpc on a side under a {\it Planck} 2015 cosmology. The corresponding mass resolution is 5.9 and $1.1\times10^7\Msun$ for dark matter and gas, respectively. Publicly available galaxy properties are tabulated at 100 snapshots.

Following the same procedure as in \S\ref{sub:data_UMachine}, we select the 1\,000 \illustris star-forming and quenched central galaxies with host halo masses closest to $\logMpeak=11.4$, 11.8, 12.3, and 12.8, respectively. Note that almost all central galaxies in high mass haloes are quenched in this simulation; consequently, selecting the 1\,000 star-forming galaxies with host halo mass closest to $\logMpeak=12.8$ would result in selecting galaxies in haloes with a broad range of masses. Motivated by this, we do not consider star-forming galaxies in such haloes. Our \illustris library thus includes 7\,000 SFHs.



\begin{table}
    \begin{center}
	\caption{\label{tab:um_properties} First and second moments of the stellar mass distribution of galaxies in haloes of different masses as predicted by \umachine and \illustris. White and grey columns show the results for star-forming and quenched galaxies, respectively.}
	\begin{tabular}{ccaca}
		\hline
		\logMpeak & \multicolumn{2}{c}{\logMstar} & \multicolumn{2}{c}{$\sigma_{\logMstar}$}\\
		\hline
		\multicolumn{5}{c}{\umachine}\\
		\hline
		11.5 &  9.50 &  9.48 & 0.23 & 0.24 \\
		12.0 & 10.33 & 10.33 & 0.24 & 0.24 \\
		12.5 & 10.76 & 10.76 & 0.20 & 0.22 \\
		13.0 & 11.03 & 11.04 & 0.20 & 0.21 \\
		\hline
		\multicolumn{5}{c}{\illustris}\\
		\hline
		11.4 &   9.27 &  9.47 &  0.24 & 0.28 \\
		11.8 &  10.18 & 10.32 &  0.16 & 0.16 \\
		12.3 &  10.71 & 10.68 &  0.15 & 0.10 \\
		12.8 &     -- & 11.08 &    -- & 0.14 \\
		\hline
	\end{tabular}
    \end{center}
\end{table}

\begin{figure}
    \begin{center}
	\includegraphics[width=\columnwidth]{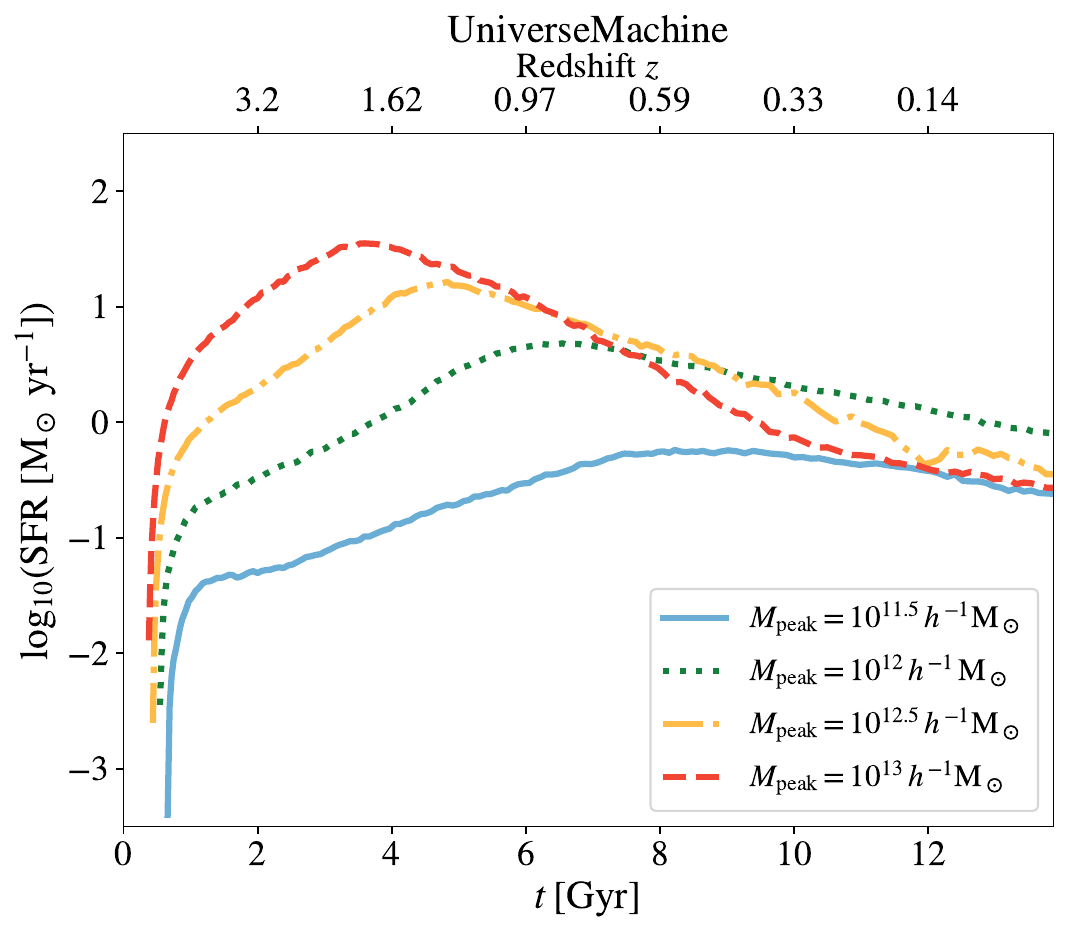}
	\includegraphics[width=\columnwidth]{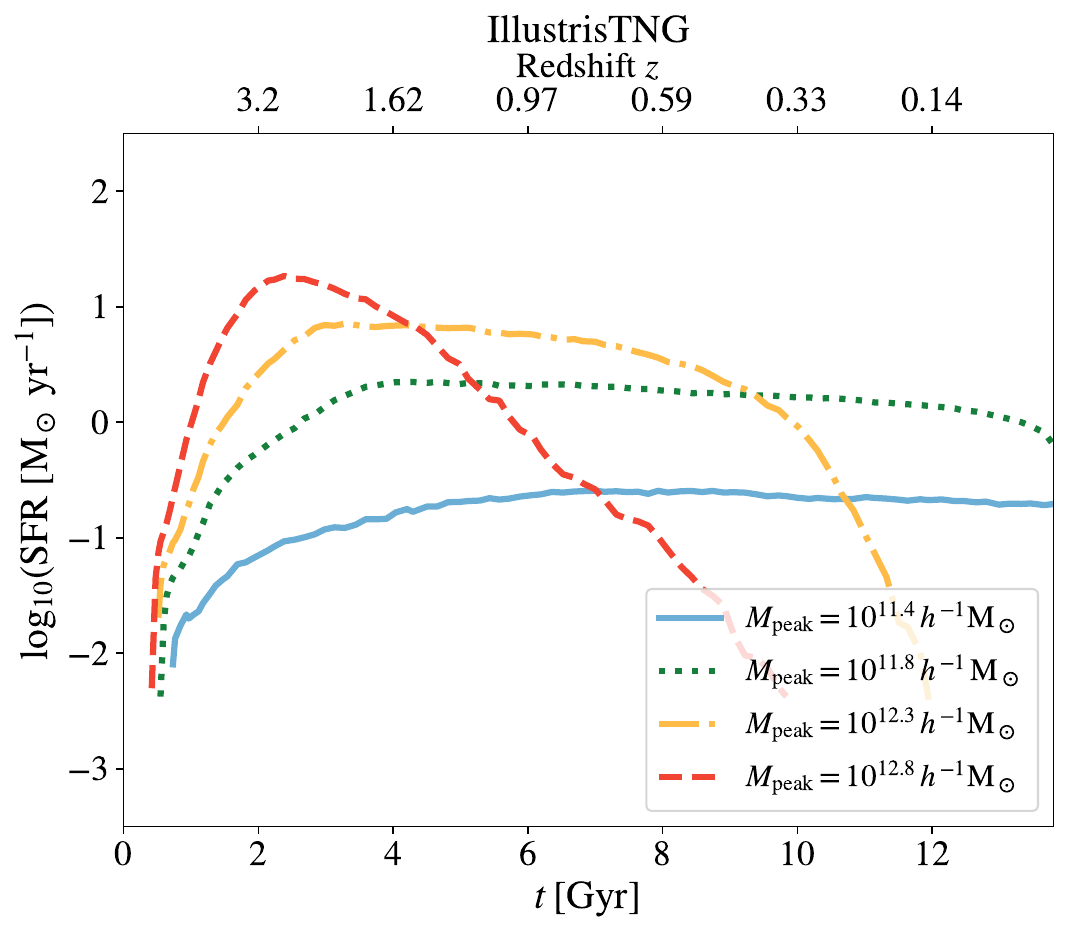}
    \end{center}
    \caption{\label{fig:SFHM_all} Median SFH of galaxies in haloes of different masses. In the top and bottom panels we display the results for \umachine and \illustris, respectively. Galaxies in more massive haloes reach the peak of their SFH earlier and present a larger stellar mass, i.e. both models predict downsizing.}
\end{figure}

\begin{figure}
    \begin{center}
	\includegraphics[width=\columnwidth]{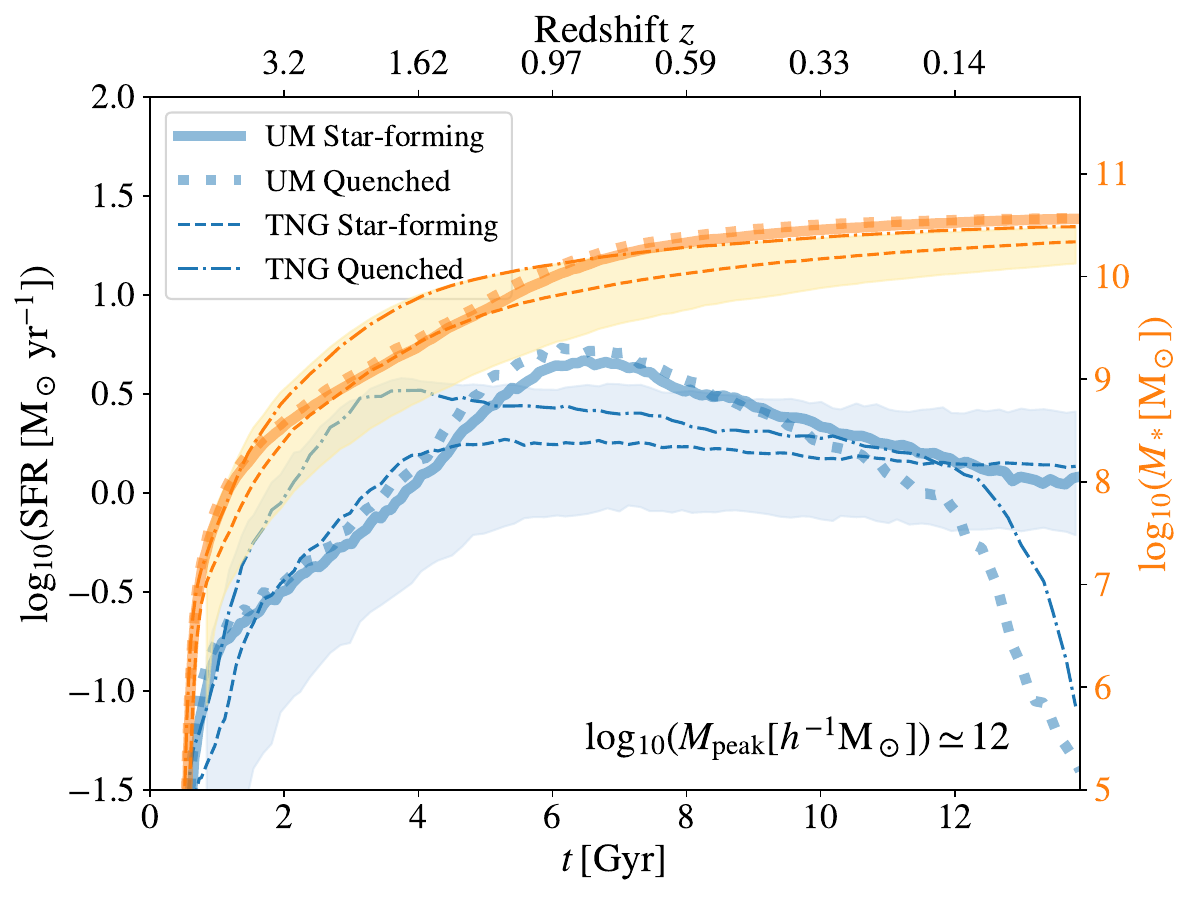}
    \end{center}
    \caption{\label{fig:SFH_Umachine_12} Median formation history of galaxies hosted by $\logMpeak\simeq12$ halos. Blue and orange lines indicate median star formation and mass histories, respectively, while solid and dotted (dashed and dot-dashed) lines show the results for star-forming and quenched \umachine (\illustris) galaxies. Shaded areas denote the region between the 16th and 84th percentiles of the results for star-forming \illustris galaxies. Star-forming and quenched galaxies present similar SFHs before $z\simeq0.1$; after this time, the SFH of quenched galaxies decreases more rapidly.}
\end{figure}

\begin{figure}
    \begin{center}
	\includegraphics[width=\columnwidth]{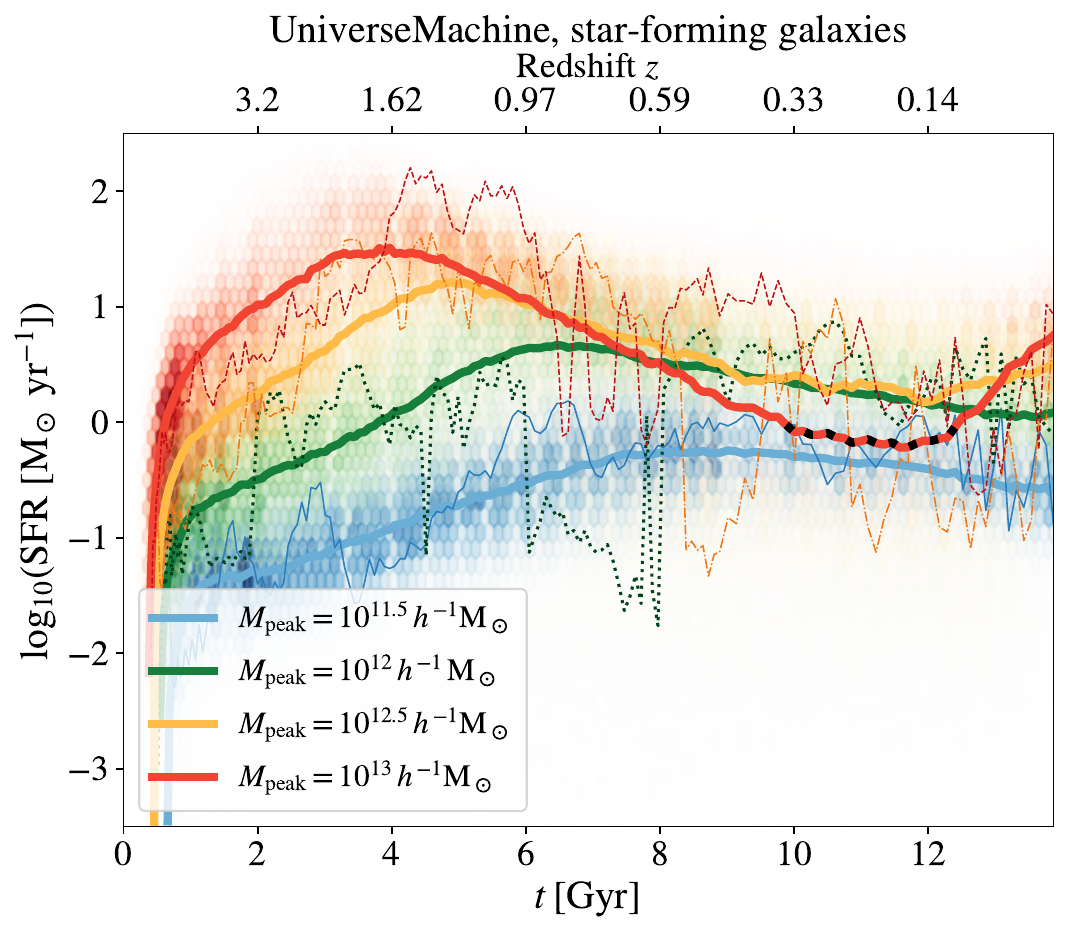}

	\includegraphics[width=\columnwidth]{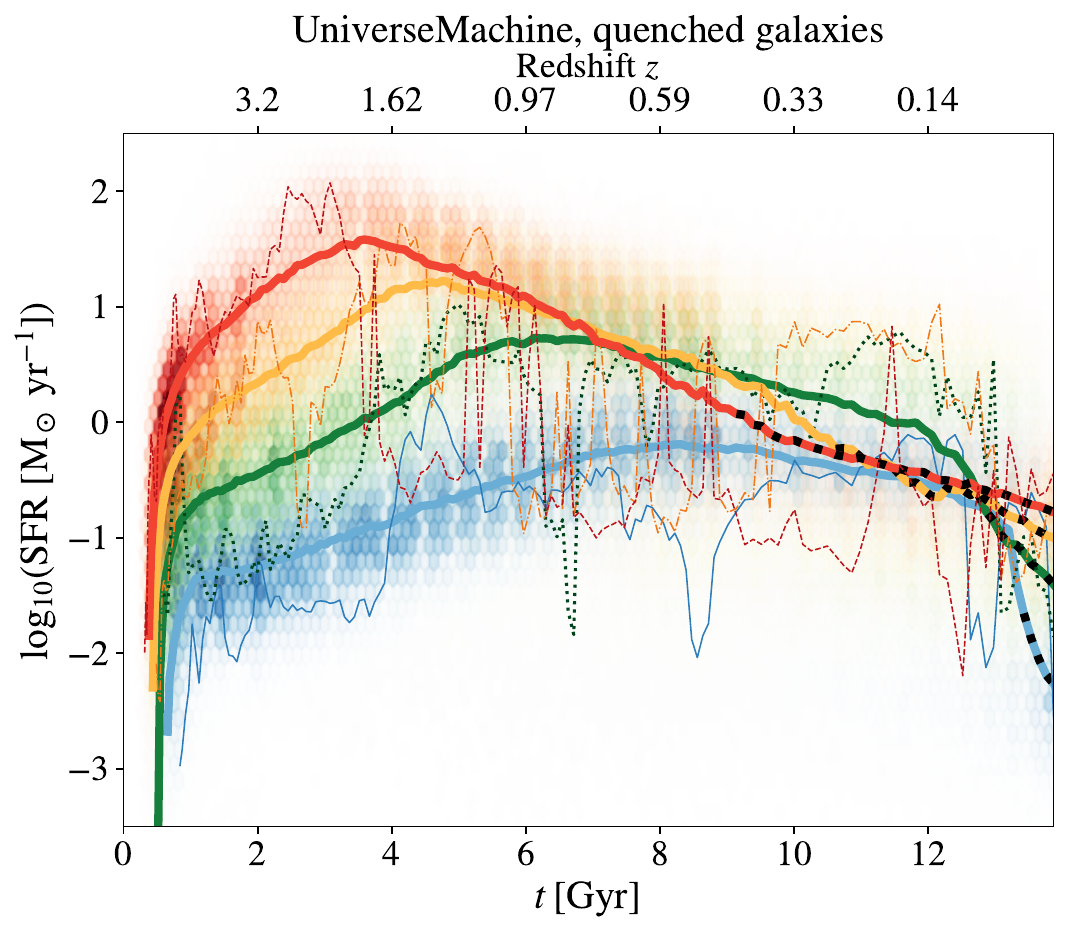}
    \end{center}

    \caption{\label{fig:SFH_Umachine} SFH of galaxies in haloes of different masses as predicted by \umachine. The top and bottom panels present the results for star-forming and quenched galaxies, respectively. Thick lines indicate median SFHs, thin lines the SFHs of randomly selected galaxies, and black stripes the periods during which galaxies would be considered quenched according to our criterion. The variety of SFHs for distinct samples is captured by two-dimensional histograms. We find that another manifestation of downsizing is that galaxies in more massive haloes become quenched at earlier times.}
\end{figure}

\section{The dependence of SFH upon host halo mass}
\label{sec:SFHhalo}

Using the \umachine and \illustris SFH libraries introduced in \S\ref{sub:data_sfh_library}, in this section we study the dependence of the SFH of star-forming and quenched central galaxies on the properties of their host haloes. As noted previously in the literature \citep{Pillepich2018b, Behroozi2019}, in both models the typical stellar mass of central galaxies increases monotonically with halo mass, and the $\Mstar$-$\Mpeak$ scaling relations exhibit good quantitative agreement with constraints from observations. For reference, in Table~\ref{tab:um_properties} we show the first and second moments of the stellar mass distribution of galaxies in different halo mass bins.

Due to the dependence of \Mstar upon \Mpeak, we naturally expect a correlation between the SFH of a galaxy and its host halo mass. To study this relationship, in Fig.~\ref{fig:SFHM_all} we present the median SFH of galaxies in haloes of different masses, showing results for \umachine and \illustris in the top and bottom panels, respectively. Each line denotes the weighted median SFH of galaxies hosted by haloes of a certain mass, as stated in the legend. We use the fraction of star-forming and quenched galaxies in haloes of the selected mass to weight the SFH of star-forming and quenched galaxies, respectively. We show weighted medians so as not to over-represent quenched (star-forming) galaxies at small (large) halo masses. In broad strokes, the SFH of galaxies in haloes of different masses look alike in \umachine and \illustris: star formation rate increases rapidly at early times, reaches a maximum several billion years ago, and decreases thereafter. Interestingly, these general characteristics show a clear dependence upon \Mpeak. The overall normalisation of the SFH at its peak increases with \Mpeak; this is sensible given that the stellar-to-halo mass relation is monotonic in both models. Furthermore, galaxies hosted by more massive haloes evolve faster, i.e. they reach the peak of their SFH at earlier times. This well-known effect is commonly known as {\em downsizing} \citep[e.g.,][]{Cowie1996}, and it results in high-mass galaxies hosting older stellar populations relative to low-mass galaxies \citep{Juneau2005, Thomas2005, Pacifici2016}.

Our criterion for classifying galaxies into star-forming and quenched only considers their specific SFR at observation time. Given that these two populations present different observational properties, it is natural to address whether their entire SFH reflects these differences. In Fig.~\ref{fig:SFH_Umachine_12} we present the median star formation and stellar mass history of galaxies hosted by $\logMpeak\simeq12$ haloes. Blue and orange lines indicate median star formation and mass histories, respectively, while solid and dotted (dashed and dot-dashed) lines show the results for star-forming and quenched galaxies as predicted by \umachine (\illustris). Shaded areas encapsulate the region between the 16th and 84th percentiles of the results for \illustris star-forming galaxies. As we can see, star-forming and quenched galaxies at $z=0$ show similar SFHs before $z\simeq0.1$, and, after this time, the SFH of quenched galaxies decreases more rapidly. Although the SFHs of both populations are noticeably different, the stellar mass of star-forming and quenched galaxies is approximately the same at all redshifts. This is consistent with the fact that galaxies in $\logMpeak\simeq12$ haloes form less than 10\% of their current stellar mass after $z\simeq0.3$, as the formation of stars peaks at higher redshift for these galaxies. Similar conclusions apply to galaxies hosted by more massive haloes, as their SFH peaks at even higher redshifts. Indeed, in Table~\ref{tab:um_properties} we show that the stellar masses of star-forming and quenched galaxies in \umachine and \illustris are compatible with one another to within one standard deviation for all halo masses studied.

To further study the dependence of SFH on halo mass for star-forming and quenched galaxies, in the top and bottom panels of Fig.~\ref{fig:SFH_Umachine} we separately display the SFHs of both populations as a function of host halo mass. This figure show the galaxy-halo connection as predicted by \umachine; see Appendix~\ref{app:tng_res} for an analogous plot for \illustris. Thick lines indicate median SFHs and thin lines the SFHs of randomly selected galaxies. Despite the smooth regularity of the median trends, we find that SFHs are very bursty, changing by orders of magnitude over short timescales. The typical amplitude of these variations is captured by the two-dimensional histograms on the background. We find that that SFHs predicted by \umachine and \illustris agree qualitatively; however, \illustris predicts a less (more) steeply decreasing SFH for star-forming (quenched) galaxies at late times, and that SFHs are more bursty in \illustris relative to \umachine. Black stripes denote the periods during which galaxies would be considered quenched according to our criterion. As we can see, in both models the lookback time at which a galaxy becomes quenched increases with its host halo mass.


\section{Influence of SFH on broad-band galaxy SED}
\label{sec:SFHSED}

In \S\ref{sec:SFHhalo} we studied the dependence of SFH on host halo mass using \umachine and \illustris, finding that both models predict that galaxies in the more massive haloes form first, reach the peak of their SFH faster, and quench earlier, i.e. galaxies in more massive haloes are more evolved. Here, we address how these trends manifest as a dependence of broad-band galaxy SEDs upon host halo mass.


\begin{figure*}
    \begin{center}
        \includegraphics[width=0.9\columnwidth]{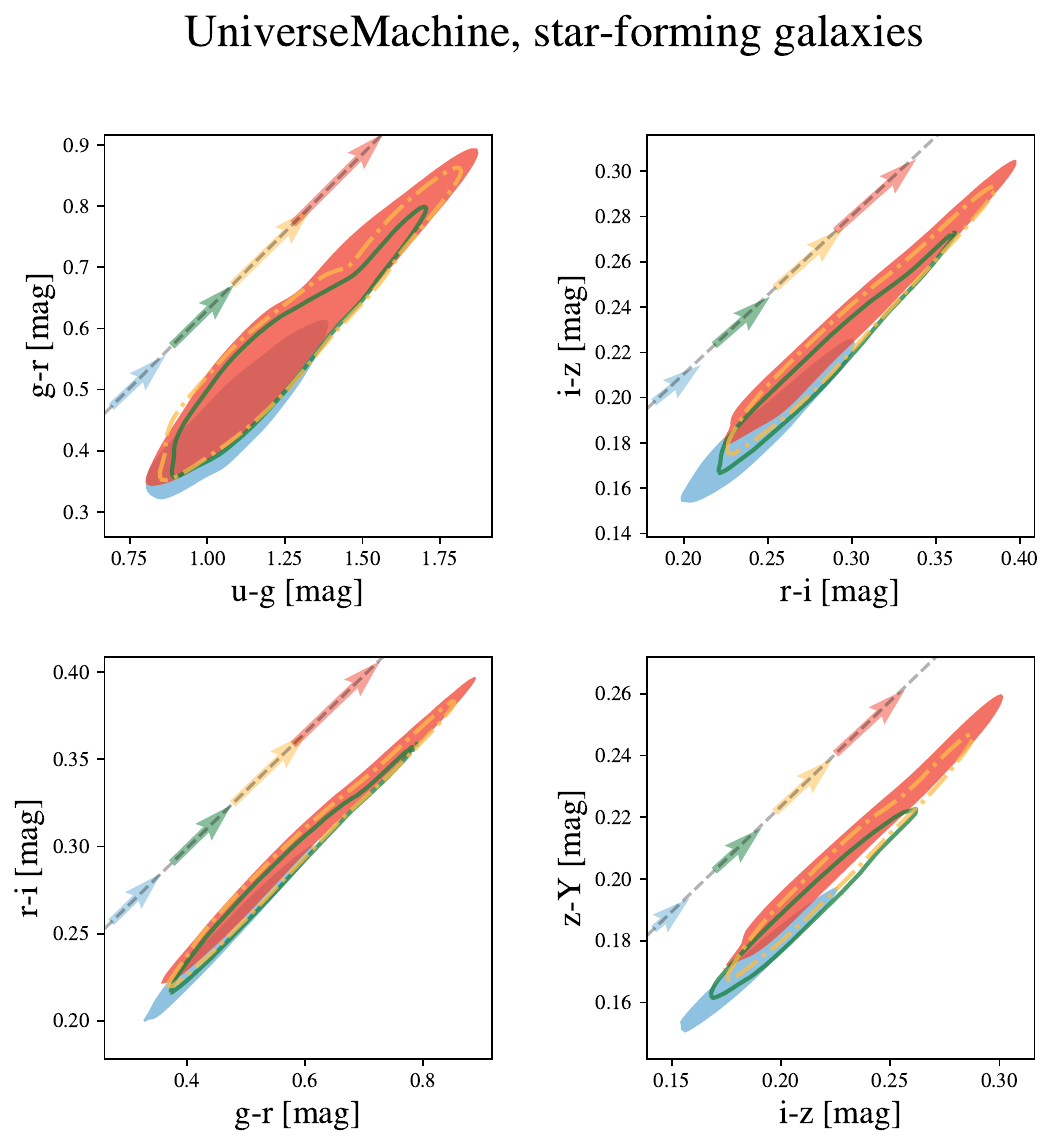}\hspace*{1cm} \includegraphics[width=0.9\columnwidth]{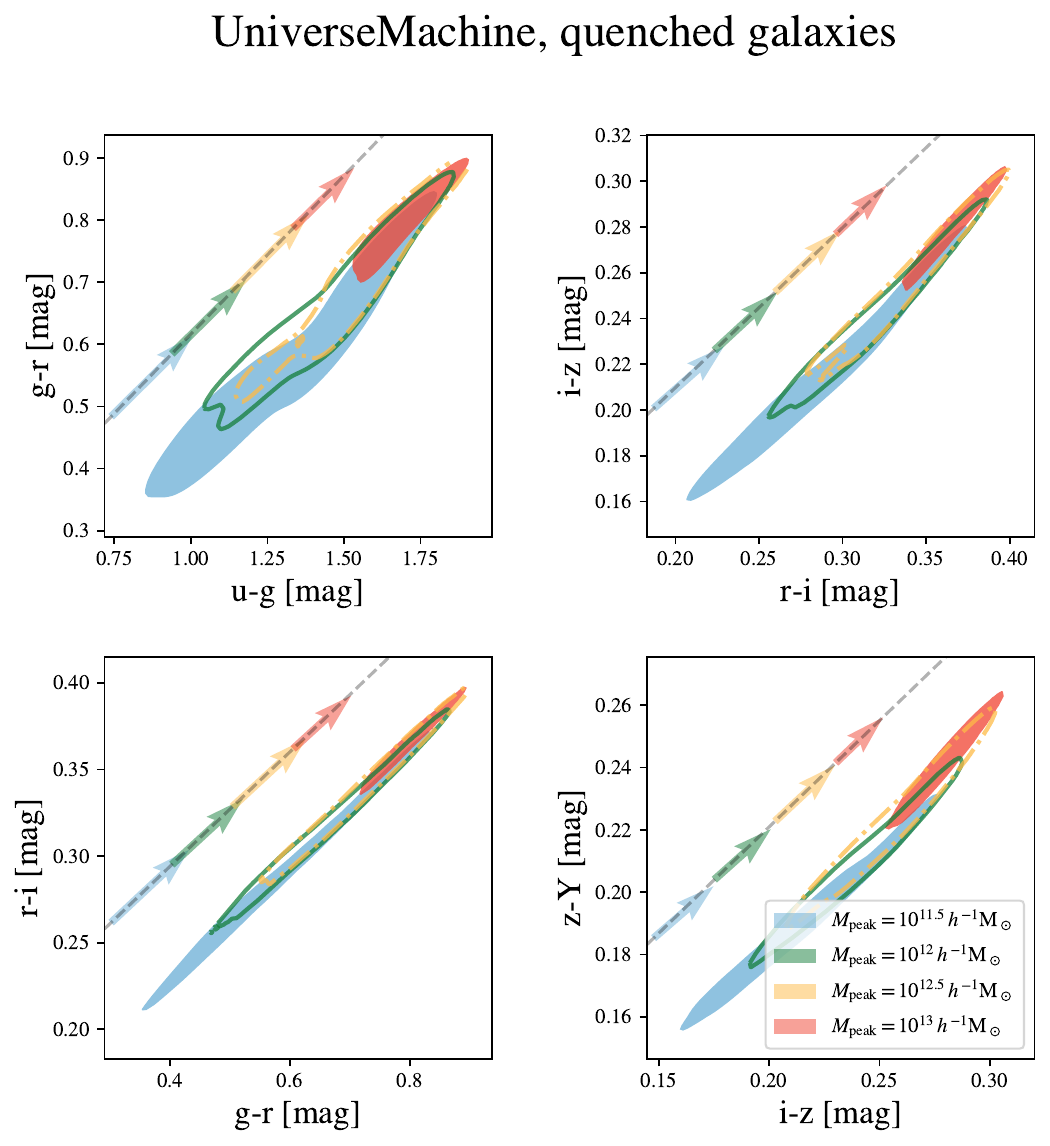}
    \end{center}
    \caption{\label{fig:SFH_colors} Influence of SFH on broad-band galaxy colours at $z=0.03$. In the left and right panels, we present the results for star-forming and quenched galaxies, respectively. We produce these colours with \galaxpy, holding fixed all galaxy properties and using the star formation histories of \umachine galaxies; thus the range colours spanned by these distributions exclusively reflects the diversity in SFH predicted by \umachine. Contours and shaded areas enclose 75\% of the distributions, arrows point in the direction of their first eigenvectors, arrow sizes indicate the standard deviation of their first PCs, and grey dashed lines denote the first eigenvector of the entire population of \umachine galaxies. Strikingly, for all galaxy populations, the grey lines run parallel to all first eigenvectors; in Appendix~\S\ref{app:tng_res} we show that the same result holds for SFH variations predicted by \illustris, demonstrating that physically-motivated SFH variations modify galaxy colours along a single direction in the five-dimensional colour space.}
\end{figure*}

\begin{table}
    \begin{center}
	\caption{\label{tab:sfh_differences} Properties of broad-band colour 
	distributions that isolate the influence of SFH variations for star-forming and quenched galaxies in haloes of different masses. We use the same coding as in Table~\ref{tab:um_properties}.}
	\begin{tabular}{ccacaca}
		\hline
		\Mpeak & \multicolumn{2}{c}{Var$^\dagger$ [\%]} & \multicolumn{2}{c}{Angle$^\ddagger$ [deg]} & \multicolumn{2}{c}{$\mathrm{VME}$ [$\times10^{-6}$ mag$^5$]}\\
		\hline
		\multicolumn{7}{c}{\umachine}\\
		\hline
		11.5 & 95.6 & 97.6 & 3.5 & 2.9 &  3.1 &  8.6\\
		12.0 & 96.5 & 97.7 & 2.4 & 2.7 &  4.9 &  5.2\\
		12.5 & 97.1 & 98.1 & 2.5 & 2.3 & 10.6 &  6.0\\
		13.0 & 97.5 & 97.8 & 1.9 & 2.8 & 15.6 &  2.0\\
		\hline
		\multicolumn{7}{c}{\illustris}\\
		\hline
		11.4 & 96.5 & 95.9 & 5.1 & 1.4 &  1.2 &  2.1\\
		11.8 & 96.3 & 97.3 & 4.4 & 1.4 &  2.1 &  3.8\\
		12.3 & 97.7 & 98.7 & 3.5 & 1.8 &  6.7 &  0.4\\
		12.8 &   -- & 98.5 &  -- & 4.7 &   -- &  0.1\\
		\hline
	\end{tabular}
	\parbox{\columnwidth}{
	 $^\dagger$ Variance explained by the first principal component. \\$^\ddagger$ Angle between the first eigenvector of the indicated sample and the SFH-direction.}
    \end{center}
\end{table}

\subsection{Influence of SFH on broad-band colours}
\label{sub:SFHSED_colors}

In this section, we explore the influence of SFH on broad-band optical and near-infrared galaxy colours from 3\,200 to 10\,800 {\AA}. In particular, we study the five colours that result from subtracting consecutive LSST bands: \{u-g, g-r, r-i, i-z, z-Y\}. 

We use \galaxpy to isolate the impact of SFH variations on broad-band colours. Specifically, we hold fixed all \galaxpy variables such as metallicity and dust properties while generating a SED for each SFH in the \umachine and \illustris libraries. Any difference between the resulting colours thus arises from SFH variations. Although these simulated colours have not been fitted to photometric observations, we will show in \S\ref{sec:colorsobv} that they broadly represent the distribution of galaxy colours from observations, thereby serving our primary purpose of capturing how physically motivated SFH variations translate into observations of galaxies in colour space.

In Fig.~\ref{fig:SFH_colors} we show the simulated colours of galaxies hosted by haloes of different masses at $z=0.03$. In the left and right panels, we present the results for star-forming and quenched \umachine galaxies, respectively; see Appendix~\ref{app:tng_res} for an equivalent figure for \illustris. At fixed \Mpeak, the scatter in colours arises from variations amongst SFHs. Even though this scatter is large, especially for star-forming galaxies, galaxies in low (high) mass halos are preferentially located closer to the bottom left (top right) of all panels. Thus, central galaxies in high mass haloes are in general redder than those in low mass haloes. We highlight that this reddening is purely due to differences in SFH, i.e. this is a direct consequence of downsizing.

For a quantitative characterisation of these colour distributions, we carry out a principal component analysis \citep[PCA,][]{Pearson1901}. PCA is an orthogonal transformation that decomposes a multivariate dataset into linearly uncorrelated variables called principal components (PC). This transformation is done in such a way that the first PC accounts for the largest possible amount of variability in the data, the second PC maximises the remaining variance under the constraint that it is orthogonal to the first PC, and so on. Therefore, PCs define an orthogonal basis in the multidimensional vector space of the data. We will refer to these vectors as eigenvectors. As succeeding components explain less and less variance, PCA can be used to reduce the dimensionality of a dataset.

PCA has been widely used in the literature in closely related contexts, such as spectral classification of galaxies \citep{Connolly1995} and quasars \citep{Yip2004}, stellar mass estimation \citep{Chen2012}, K-corrections \citep{Blanton2007}, and inferring spectra from optical imaging \citep{kalmbach_connolly17}. In this work, the application of PCA to galaxy colours is motivated by the obvious correlation between broad-band colours shown in Fig.~\ref{fig:SFH_colors}. In Table~\ref{tab:sfh_differences} we show the variance explained by the first PC of the simulated colours of \umachine and \illustris star-forming and quenched galaxies in different halo mass bins. Independently of population and host halo mass, we find that the first PC of each distribution explains more than 96\% of the total variance in colours, implying that most of the information about SFH encoded in broad-band photometry can be captured by just a single linear combination of galaxy colours. This conclusion holds for both \umachine and \illustris, even though these two models have radically distinct implementations of the galaxy-halo connection and predict SFHs with different shapes and levels of burstiness.

The previous result suggests the existence of a single direction in colour space that corresponds to physically-motivated variations in star formation history. It is natural to consider whether this direction varies across different populations, halo masses, or SFH models. To address this question, we repeat the above PC decomposition, but for our dataset we use the entire collection of \umachine (or \illustris) galaxies, without separating them into quenched/star-forming subpopulations or into different halo mass bins. We find that the first PCs of both \umachine and \illustris simulated colours explain more than 98\% of the data variance, and moreover that there is an angle of approximately 0.5 degrees between their first eigenvectors. In what follows, we refer to the orientation of the first PC constructed in this fashion as {\em the SFH-direction}.

We now compare the SFH-direction to the first eigenvector resulting from the PC decomposition of the colours of star-forming and quenched galaxies residing in halos of different masses. In Table~\ref{tab:sfh_differences} we present the angle between the SFH-direction and each of these first eigenvectors. We find that this angle is smaller than 5 degrees for all subsamples, and thus SFH variations modify colours in the same direction of the colour space for all populations and halo masses. We illustrate this result in Fig.~\ref{fig:SFH_colors}; in this figure dashed lines indicate the SFH-direction, arrows point in the direction of the first eigenvector of each subpopulation, and the sizes of the vectors denote the standard deviation of colours projected onto this direction. In accord with the invariance of the angles in Table~\ref{tab:sfh_differences}, all eigenvectors are well-aligned in all projections of the colour space. We therefore conclude that {\em physically-motivated SFH variations result in galaxy colours moving along a single direction in the five-dimensional colour--colour space.}

In Fig.~\ref{fig:SFH_colors} we can also see that the scatter in galaxy colours depends strongly on \Mpeak. We quantify this scatter by computing the volume of the minimum ellipsoid (VME) enclosing $68\,\%$ of each colour distribution. To do so, we start by projecting all distributions onto their eigenvectors. Then, we compute the VME using the following expression

\begin{equation}
\mathrm{VME} = \frac{\upi^{n/2}\,[\Phi_n(\alpha)]^n} {\Gamma (1 + n/2)} \prod_{i=1}^n \sigma_i,
\end{equation}

\noindent where $n=5$ is the number of colours, $\Gamma$ refers to the Gamma function, $\Phi_n$ is the quantile function of a $\chi^2$ distribution with $n$ degrees of freedom, $\alpha=0.68$ indicates the percentage of the distribution that the ellipsoid encloses, and $\sigma_i$ is half the difference between the 16th and 84th percentiles of projections onto the $i$th PC. 

In Table~\ref{tab:sfh_differences} we present the VME of star-forming and quenched galaxies hosted by halos of different masses. Overall, the VME of the colours of star-forming (quenched) galaxies increases (decreases) with \Mpeak in both \umachine and \illustris. In the next section, we proceed to explore the origin of these trends.


\begin{figure}
	\includegraphics[width=\columnwidth]{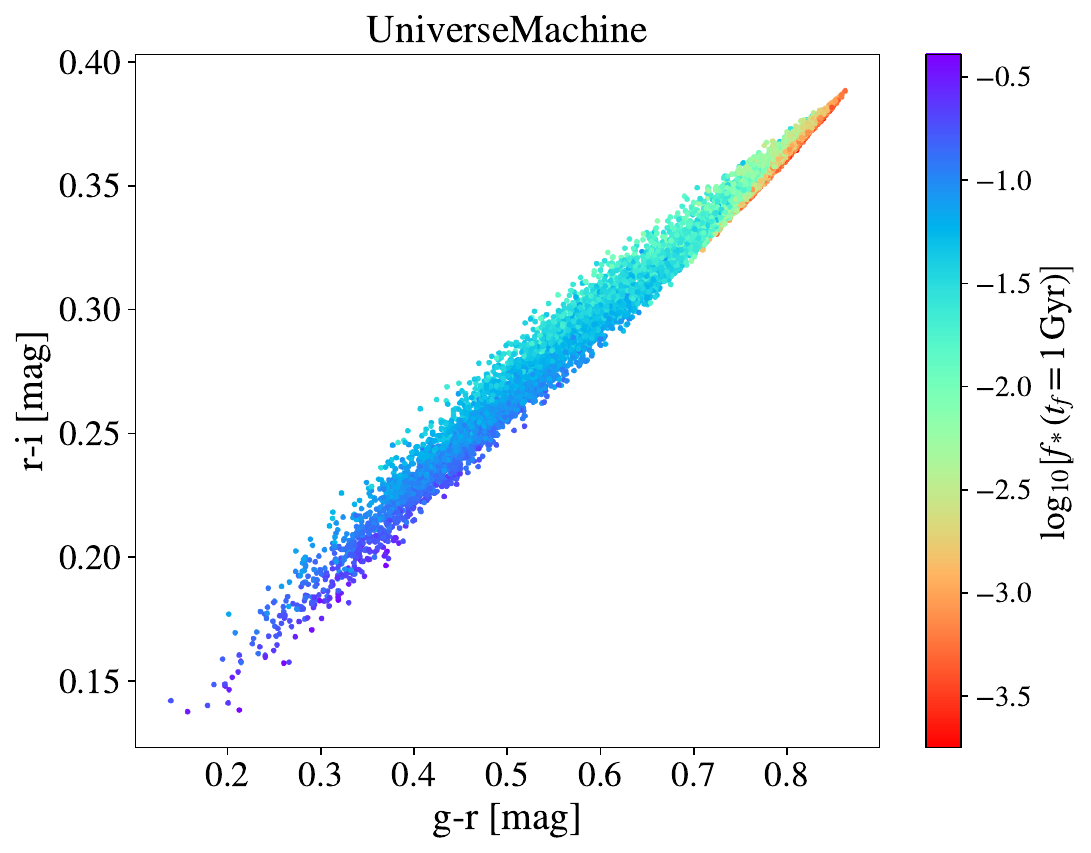}
    \caption{\label{fig:colordif_IM} Correlation between \{g-r, r-i\} colours and \fstar, the fraction of stellar mass formed over the last billion years. Each point represents a \umachine galaxy at $z=0.03$. As expected, galaxies with larger values of \fstar are bluer. Irrespective of halo mass, the correlation between \fstar and galaxy colours is exceptionally tight, and exhibits little-to-no dependence upon galaxy assembly history prior to the last billion years.}
\end{figure}

\begin{figure}
	\includegraphics[width=\columnwidth]{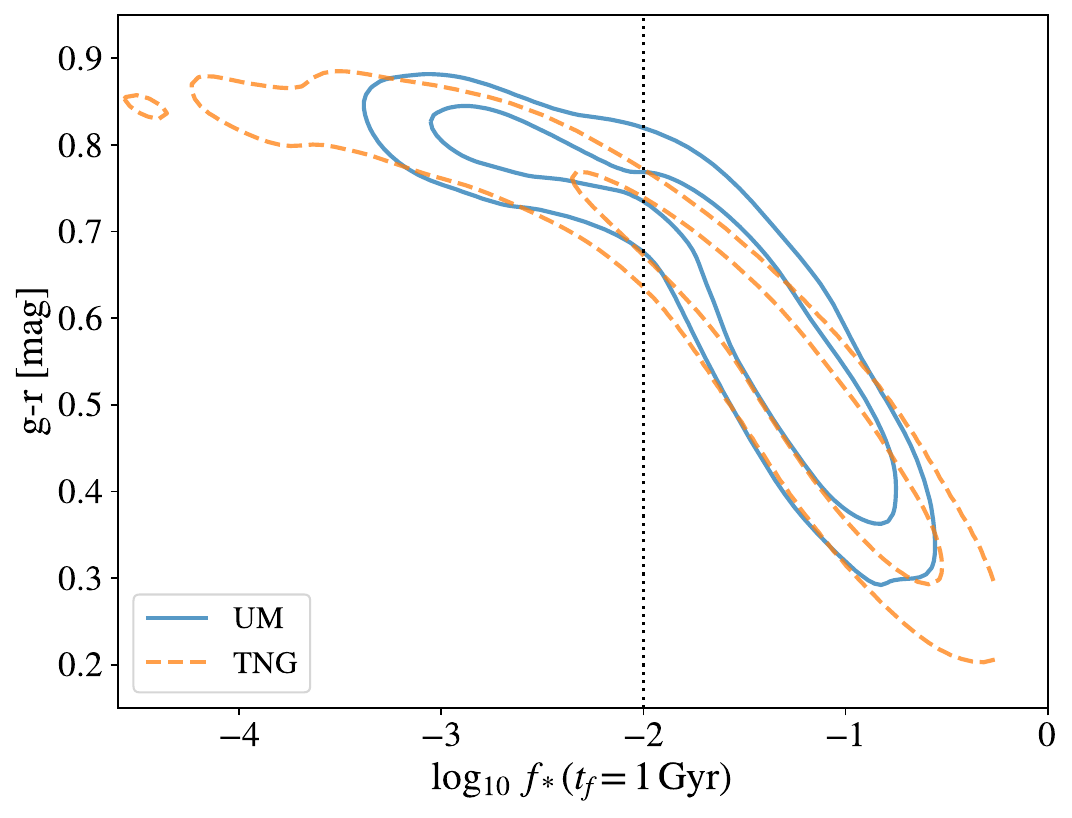}
	\includegraphics[width=\columnwidth]{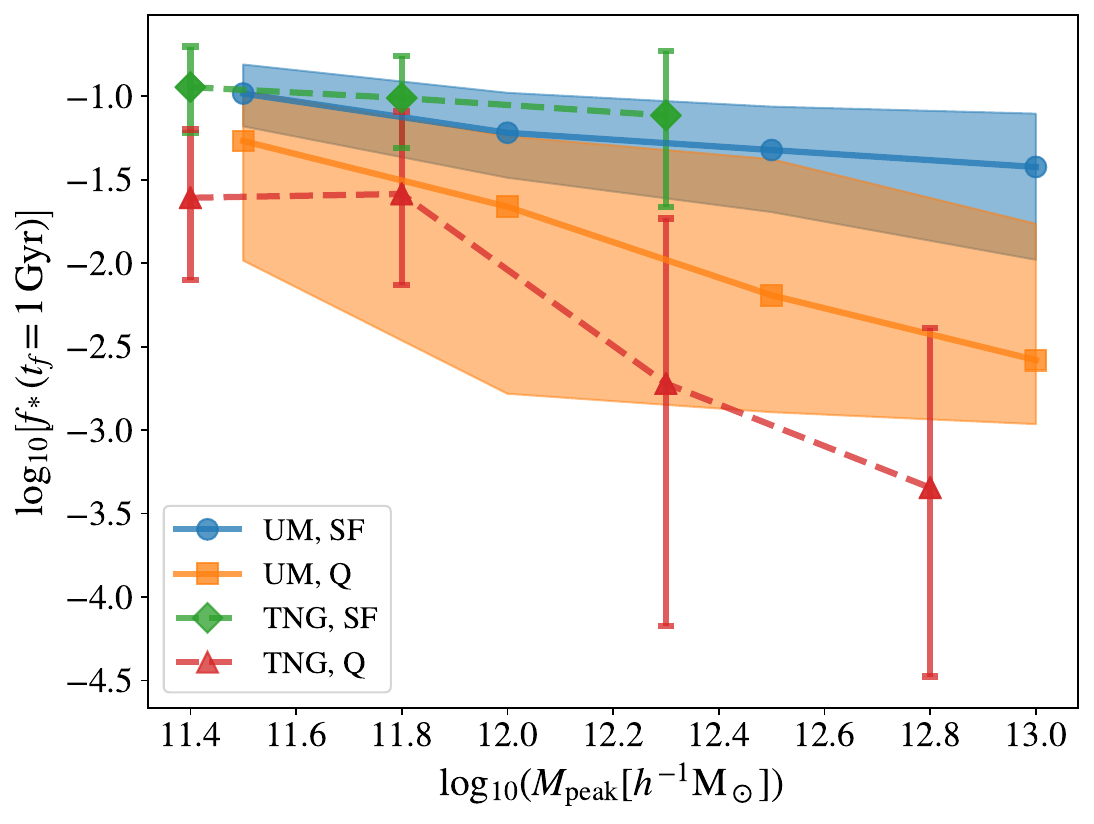}
    \caption{\label{fig:fstar_mpeak} Origin of the correlation between galaxy colours and SFH. {\bf Top panel.} Dependence of the g-r colour on \fstar at $z=0.03$. Blue solid and orange dashed contours indicate the results for \umachine and \illustris, respectively, and the vertical dotted line indicates $\fstar=0.01$. As we can see, there is a strong relation between \fstar and g-r: for $\fstar>0.01$ galaxies get rapidly bluer as \fstar increases, while for $\fstar<0.01$ galaxies practically present the same red colours. This result is also valid for other broad-band colours. {\bf Bottom panel.} Distribution of \fstar for galaxies hosted by haloes of different masses. Error bars and shaded areas indicate 16th and 84th percentiles. As expected, the average value of \fstar decreases as \Mpeak increases. The top and bottom panels together provide a simple way to understand the tightness of the red sequence.}
\end{figure}

\subsection{Physical origin of colour variations owing to SFH differences}
\label{sub:SFHSED_origin_colors}

After a star formation episode, a galaxy forms stars according to its initial mass function, and, if this episode is sufficiently strong, newly formed O and B stars quickly dominate optical and near-infrared wavelengths \citep[e.g.,][]{Renzini1986, Charlot1991}. However, O and B stars do not dominate these wavelengths for more than a few million years if no further formation of stars takes place; this is because the time a star spends in the main sequence (MS) decreases rapidly with its mass. Massive, blue stars thus leave the main sequence quickly, and lower mass, redder stars progressively dominate short optical wavelengths. This picture is nonetheless more complicated for longer optical and near-infrared wavelengths; this is the result of horizontal, asymptotic giant, and red giant branch stars contributing substantially to these wavelengths 0.1, 1, and 10 billion years after a star-forming episode \citep[e.g.,][]{Charlot1991, Maraston1998}, respectively. These two processes explain why the optical and NIR colours of quenched galaxies become progressively redder.

The situation explained above is different for a galaxy presenting further formation of stars; this is because such galaxy would present a steady supply of massive, blue MS stars, which would dominate short optical wavelengths \citep[e.g.,][]{Bruzual1993}. For longer wavelengths, we face a more complicated scenario: for some billion years blue MS stars would dominate the light of this galaxy but progressively, as more MS stars evolve into the horizontal, asymptotic giant, and red giant branch, the light of this galaxy would get increasingly redder \citep[e.g.,][]{Bruzual1993}. In summary, the optical and near-infrared colours of a galaxy undergoing an intense star-forming episode become extremely blue because massive MS stars dominate these wavelengths, while those of a galaxy no experimenting any formation of stars for a few billion years get red due to post MS stars dominating these. This scenario is obviously more complex for galaxies with intermediate star formation, and it is metallicity dependent; as the metallicity of a galaxy increases, so it does the dominance of its light by MS stars \citep[e.g.,]{Maraston2005}. Based on this intuitive picture, we proceed to investigate why galaxies with different SFHs have distinct colours.

This aforementioned scenario motives exploration of the following proxy for galaxy colours: the fraction of stellar mass formed over the last $t_f$ years before observation time $t_0$, $f_*(t_f) \equiv \frac{M_*(t_0) - M_*(t_0-t_f)}{M_*(t_0)}$. In Fig.~\ref{fig:colordif_IM}, we display the projection of the simulated colours produced in \S\ref{sub:SFHSED_colors} for \umachine onto the {g-r; r-i} plane (see Appendix~\ref{app:tng_res} for an analogous figure for \illustris). Symbols are colour-coded according to \fstar and indicate the results for individual galaxies. We find that the g-r and r-i colours present a very strong correlation with \fstar for star-forming and quenched galaxies in haloes of different masses, where these colours get bluer as \fstar increases. Of course, this correlation is entirely expected based on the well-established physical model introduced above; our concern in the remainder of this section is a quantitative assessment of the explanatory power of \fstar when applied to the physically-motivated star formation histories predicted by \illustris and \umachine.

We further test the accuracy of $f_*$ as a proxy for SFH variations in galaxy colours by computing the correlation between the projection of the colours shown in Fig.~\ref{fig:colordif_IM} onto the SFH-direction, exploring $f_*(t_f)$ for $t_f=0.25$, 0.5, 1, 1.5, and 2 Gyr. We find that the strength of this correlation at $t_f<1\,\mathrm{Gyr}$ exceeds $r_s=0.86$ and 0.96 for \umachine and \illustris, respectively, with very little dependence upon $t_f$. In contrast, this correlation weakens rapidly for timescales larger than $t_f\gtrsim1\,\mathrm{Gyr}$. This characteristic timescale reflects the time at which all stars more massive than $2\,\Msun$ have left the main sequence. We note that this timescale is wavelength dependent; young (old) stars are much brighter on wavelengths shorter (longer) than those studied here \citep[e.g.,][]{Maraston1998}, and thus our results only hold for optical and near-infrared rest-frame colours.

For a better understanding of the relation between galaxy colours and \fstar, in the top panel of Fig.~\ref{fig:fstar_mpeak} we display simulated g-r colours as a function of \fstar. Blue solid and orange dashed contours indicate the results for \umachine and \illustris, respectively, where the inner and outer contour enclose 60 and 90\% of the corresponding colour distribution. Generally speaking, the g-r colour gets redder as \fstar decreases, a result that also applies to other optical colours. We can readily see that there is a strong dependence of g-r on \fstar for galaxies with $\fstar\gtrsim0.01$: g-r increases by $\sim0.4$ mag between $\fstar=0.1$ and 0.01. On the other hand, the dependence of g-r upon $\fstar$ rapidly weakens for galaxies with $\fstar<0.01$, so that all such galaxies have nearly the same red colours. This trend can again be understood in terms of the basic physical picture outlined above: galaxies with $\fstar<0.01$ form essentially no new blue stars over the last billion years, and so their stellar light is dominated by old main sequence stars and asymptotic and red giant branch stars. Note that at fixed \fstar colours produced using \umachine and \illustris SFHs are very similar. This is remarkable, given the differences between SFHs predicted by both models. We further study this result in a companion paper (Chaves-Montero et al. in prep.). 

Given the correlation between SFH and halo mass, we also expect a similar connection between \fstar and \Mpeak. In the bottom panel of Fig.~\ref{fig:fstar_mpeak}, we show the distribution of \fstar for galaxies hosted by haloes of different masses. Blue and orange (green and red) colours indicate the results for star-forming and quenched \umachine (\illustris) galaxies, respectively. Symbols denote medians, while error bars show 16th and 84th percentiles. Overall, star-forming galaxies present a larger value of \fstar relative to quenched galaxies, and for both populations \fstar decreases as \Mpeak increases. These basic scaling relations with \fstar drive the trend of star-forming (quenched) galaxies having preferentially blue (red) colours, as well as the tendency of central galaxies to be redder in haloes of increasing mass. Note that the results are very similar for \umachine and \illustris.

We can gain further insight into the distribution of galaxy colours by consideration of how \fstar varies with halo mass. The collection of star-forming and quenched galaxies in haloes less massive than $\logMpeak<12.2$ tend to have a broad range of values of \fstar that exceed 0.01. As shown in the top panel of Fig.~\ref{fig:fstar_mpeak}, the g-r colour has a strong dependence upon \fstar in this regime, and so the wide range of values of $\fstar>0.01$ naturally gives rise to the wide range of colours of galaxies in low-mass haloes. In \S\ref{sub:SFHSED_colors}, we also observed that the scatter in the colours of star-forming galaxies grows with \Mpeak; this is consistent with the broadening of the \fstar distribution as \Mpeak increases for this population (see the blue and green curves in the bottom panel of Fig.~\ref{fig:fstar_mpeak}). On the other hand, for quenched galaxies in massive haloes, the distribution of \fstar rapidly plummets below $0.01$ with increasing halo mass; accordingly, quenched galaxies in massive halos have a much more restricted range of predominantly red colours. These results provide a simple way to understand the tightness of the red sequence.


\begin{figure}
	\includegraphics[width=\columnwidth]{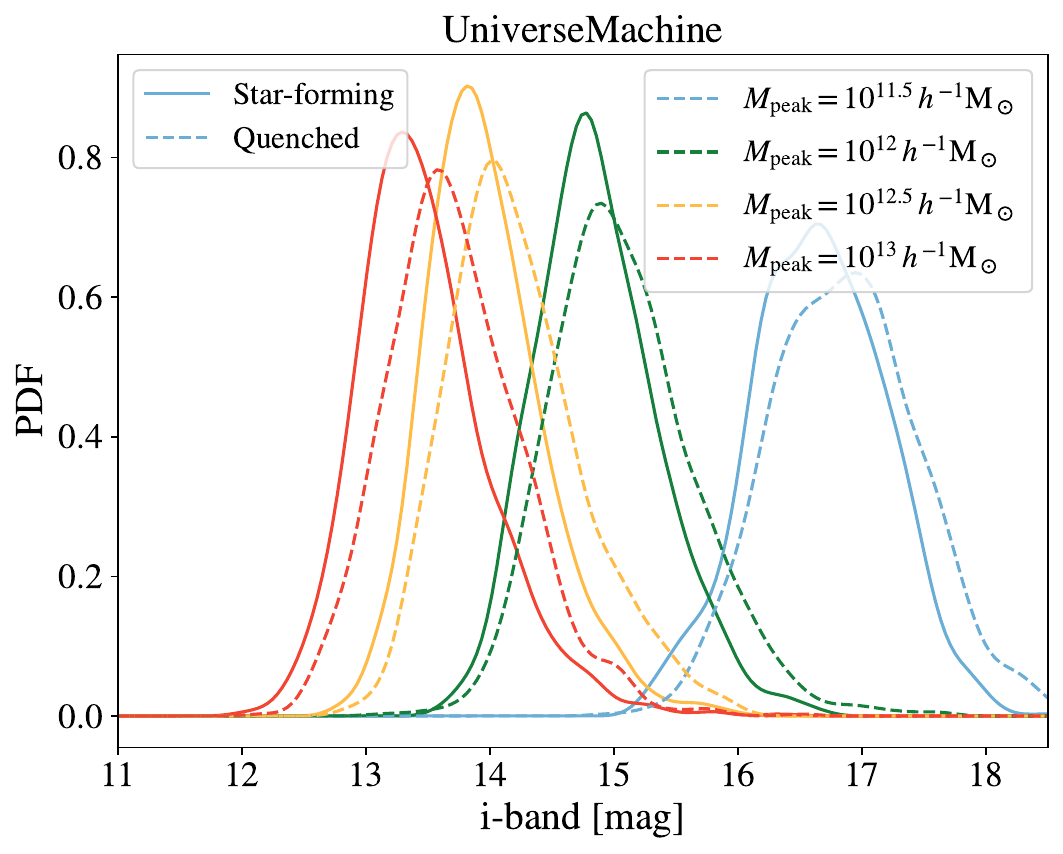}
	\includegraphics[width=\columnwidth]{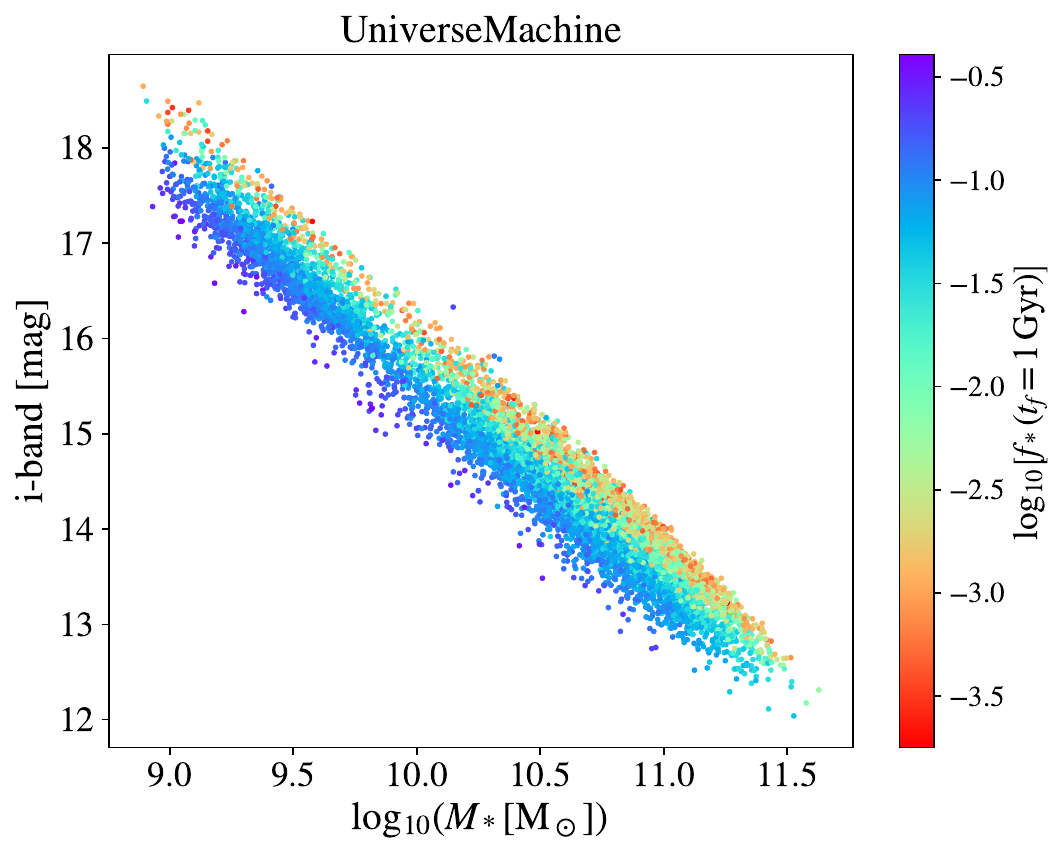}
    \caption{\label{fig:Mpeak_iband} Dependence of galaxy luminosity on SFH. {\bf Top panel}. Distribution of i-band magnitudes for galaxies in haloes of different masses at $z=0.03$. We generate these luminosities using SFHs predicted by \umachine. Solid and dashed lines present the results for star-forming and quenched galaxies, respectively. The luminosity of a galaxy increases with \Mpeak, and at fixed \Mpeak star-forming galaxies are slightly brighter than quenched galaxies. {\bf Bottom panel}. Observed i-band magnitudes of individual galaxies as a function of their stellar masses. Symbols are colour-coded according to \fstar. Even though luminosities and stellar masses are strongly correlated, this relation presents a small scatter that is mostly captured by \fstar.}
\end{figure}

\subsection{Impact of SFH on observed magnitudes}
\label{sub:SFHSED_mag}

The stellar mass of a galaxy is strongly correlated with its luminosity. This relation is especially tight in the near-infrared, as near solar-mass stars dominate the near-infrared luminosity of a galaxy as well as its stellar mass \citep{Gardner1997}. Given that more massive galaxies live in more massive haloes, we also expect a strong correlation between near-infrared luminosity and \Mpeak. In the top panel of Fig.~\ref{fig:Mpeak_iband}, we display the distribution of i-band magnitudes for galaxies in haloes of different masses at $z=0.03$ as predicted by \umachine. Solid and dashed lines denote the results for star-forming and quenched galaxies, respectively. As expected, galaxies in more massive haloes are in general brighter, albeit with significant scatter. Interestingly, we find that star-forming galaxies are slightly brighter than quenched galaxies at fixed halo mass. This trend is explained by a) star-forming galaxies having more blue stars relative to quenched galaxies, and b) the light-to-mass ratio of blue stars is larger than that of red stars. The corollary of this tendency is that at fixed \Mstar, the luminosity of a galaxy depends on its colours. This is a well-known result that has been used to improve the precision of stellar masses derived from photometric observations \citep[e.g.,][]{Bell2003}. The relation between luminosity and colours motivates the principal concern of this section, which is to study the dependence of galaxy luminosities upon \fstar, our proxy for galaxy colours.

In the bottom panel of Fig.~\ref{fig:Mpeak_iband}, we display the relation between observed i-band magnitudes and \Mstar for \umachine (in Appendix~\S\ref{app:tng_res} we present an analogous figure for \illustris). Symbols are colour-coded according to \fstar and indicate the results for individual galaxies. We find that the Spearman rank correlation coefficient between i-band magnitudes and stellar masses is -0.97. Even though this correlation is very strong, we can see that there is still some scatter between i-band magnitudes and stellar masses. Remarkably, this scatter is mostly captured by \fstar: at fixed \Mstar the brightest (dimmest) galaxies are those with the largest (smallest) \fstar. More quantitatively, we estimate the fraction of the scatter captured by \fstar by fitting two different Gaussian processes to the data; the first (second) considers \Mstar (\Mstar and \fstar) as dependent variable(s) to predict i-band magnitudes. We find that the standard deviation of the residual between data and predictions is a factor of two smaller after considering \fstar in addition to \Mstar.

Assembling the basic trends covered in this and the previous section, we have now a full picture of the influence of SFH on broad-band galaxy SEDs: galaxies get brighter as their stellar mass increases and, at fixed \Mstar, galaxies with a larger value of \fstar present bluer colours and slightly larger luminosities. Furthermore, the fraction of quenched galaxies increases with host halo mass as a result of the increasing abundance of galaxies with small values of \fstar at high masses.


\begin{figure}
	\includegraphics[width=\columnwidth]{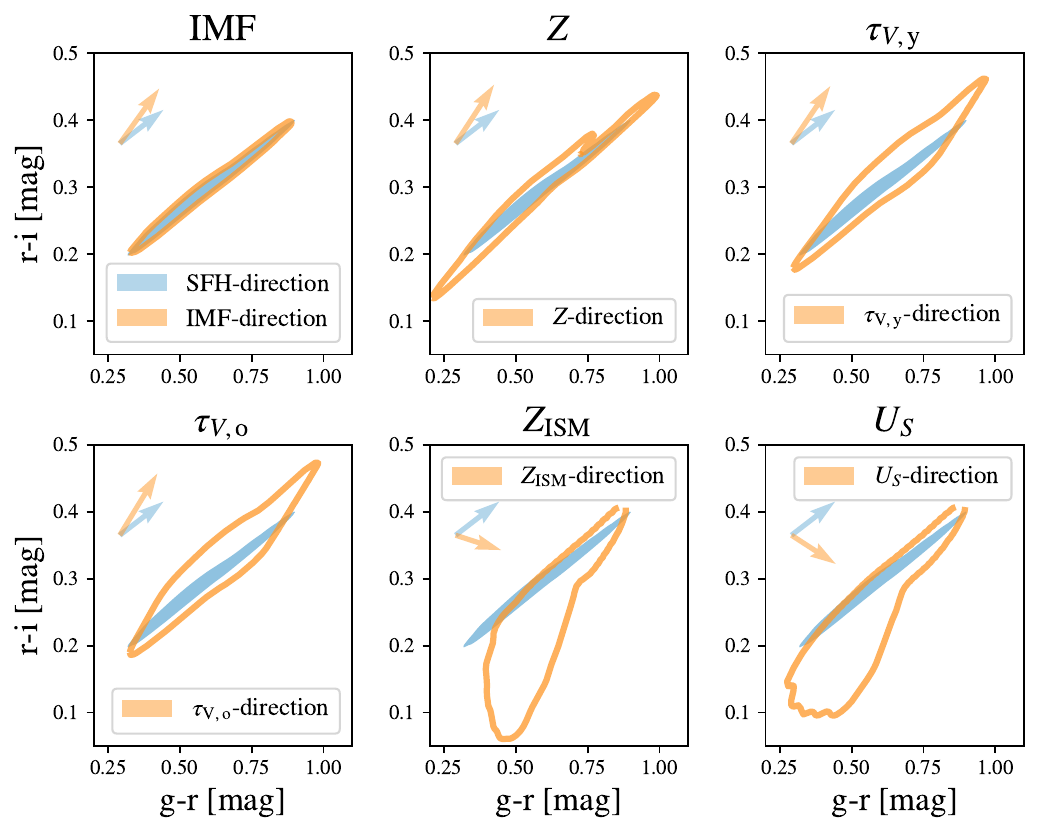}
    \caption{\label{fig:diff_params} Influence of IMF, metallicity, dust attenuation, and nebular emission lines on broad-band colours at $z=0.03$. In each panel, orange contours show the impact of the \galaxpy parameter indicated at the top of the panel, and blue shaded areas show the impact of SFH. Orange and blue arrows denote the directions in colour space associated with variations in the target galaxy property and  SFH, respectively. As we can see, changes in metallicity, dust attenuation, and nebular emission modify colours in distinct directions of the colour space relative to SFH variations.}
\end{figure}

\begin{table}
    \begin{center}
	\caption{\label{tab:extra_params} Impact of IMF, metallicity, dust emission, and nebular emission lines on broad-band colours. Results are displayed in the format $50\mathrm{th}^{75\mathrm{th}-50\mathrm{th}}_{25\mathrm{th}-50\mathrm{th}}$.}
	\begin{tabular}{ccccc}
		\hline
		Param. & Var.$^*$[\%] & Ang. SFH-dir$^\dagger$ [$^{\circ}$] & Imp. SFH-dir$^\ddagger$ [$^{\circ}$]\\
		\hline
		IMF                 & $99.79^{+0.06}_{-0.70}$ & $21^{+27}_{-7}$  & $0.06^{+0.28}_{-0.03}$ \\
		$Z$                 & $89.7^{+1.2}_{-1.3}$    & $35^{+16}_{-17}$ & $5.9^{+5.4}_{-1.4}$ \\
		$\tau_{V,{\rm y}}$  & $99.86^{+0.12}_{-0.05}$ & $12^{+5}_{-3}$   & $0.70^{+0.06}_{-0.25}$ \\
		$\tau_{V,{\rm o}}$  & $99.89^{+0.10}_{-0.44}$ & $27^{+10}_{-8}$  & $0.8^{+0.7}_{-0.3}$ \\
		$Z_{\rm ISM}$       & $97.6^{+1.5}_{-2.2}$    & $22^{+2}_{-1}$   & $8.9^{+0.6}_{-0.9}$ \\
		$U_S$               & $97.3^{+1.1}_{-1.3}$    & $55^{+4}_{-6}$   & $9.0^{+0.8}_{-1.0}$ \\
		\hline
	\end{tabular}
	\parbox{\columnwidth}{
	$^*$~Variance explained by the first PC of the indicated parameter.\\
	$^\dagger$~Angle between the direction in colour space associated with variations in the indicated parameter and the SFH-direction.\\
	$^\ddagger$~Impact of variations in the indicated parameter on the SFH-direction.}
    \end{center}
\end{table}

\section{Influence of galaxy properties on broad-band colours}
\label{sec:colorsgprop}

In \S\ref{sub:SFHSED_colors} we studied the influence of SFH on broad-band colours, finding that physically-motivated SFH variations modify galaxy colours along a single direction in colour space. In this section we address the impact of variations in IMF, metallicity, dust attenuation, and nebular emission line ratios on broad-band colours. 

We investigate the influence of galaxy properties on colours using a similar PCA-based approach as in \S\ref{sub:SFHSED_colors}. We will explore the following galaxy features:

\begin{itemize}
    \item {\em Initial Mass Function.} The fiducial configuration of \galaxpy assumes a Chabrier IMF. To study the dependence of galaxy colours on this property, we also explore Kroupa and Salpeter IMFs.
    
    \item {\em Metallicity.} In \S\ref{sub:SFHSED_colors} we generated colours presuming a solar metallicity. In what follows, we also produce colours for four sub-solar and one super-solar metallicity: $Z=0.0001$, 0.0004, 0.004, 0.008, and 0.05.
    
    \item {\em Dust attenuation.} We explore the influence of each of the two parameters controlling the \galaxpy dust attenuation model separately. We sample each of these two parameters using ten linearly-spaced values spanning the intervals $\tau_{V,\mathrm{y}} \in [0,3]$ and $\tau_{V,\mathrm{o}} \in [0,1]$.
    
    \item {\em Nebular emission lines.} The fiducial configuration of \galaxpy does not consider nebular emission lines at all. In \S\ref{sub:data_galaxpy} we presented the \galaxpy emission line model, which includes two free parameters: $Z_{\rm ISM}$ and $U_S$. We explore the following values: $Z_{\rm ISM}=0.0001$, 0.0002, 0.0005, 0.001, 0.002, 0.004, 0.006, 0.008, 0.010, 0.014, 0.017, 0.020, 0.030, 0.040 and $\log_{10}U_S=-1$, -1.5, -2, -2.5, -3, -3.5, -4.
\end{itemize}

To study the impact of a certain \galaxpy parameter on colours, we first select a single galaxy history from one of the SFH libraries introduced in \S\ref{sub:data_sfh_library}. We then produce colours for every value of the \galaxpy parameter under consideration, while holding fixed all other \galaxpy parameters to their fiducial values. After that, we iterate over every SFH in the libraries for each target \galaxpy parameter. We thus end up with a new set of simulated colours for each value of the \galaxpy parameters studied, allowing us to isolate the influence of the corresponding physical effects on colours.

In Fig.~\ref{fig:diff_params} we show the impact of IMF, metallicity, dust attenuation, and nebular emission lines on galaxy colours in the \{g-r, r-i\} plane at $z=0.03$. Orange contour indicate the distribution of colours resulting from varying the \galaxpy parameter indicated by the title at the top of each panel; for reference, a blue shaded area displays the distribution of colours arising only from SFH variations. Both contours and shaded areas contain 90\% of the distributions. Broadly speaking, when varying metallicity, dust attenuation, and nebular emission line ratios, the area of colour space spanned by galaxy colours is significantly expanded beyond the extent of that resulting from SFH variations alone, in accord with previous results \citep[e.g.,][]{Pacifici2015}.

Fig.~\ref{fig:diff_params} suggests that variations in SFH have a qualitatively different influence on galaxy colours relative to other galaxy properties. To quantitatively compare the impact of the SFH and that of other galaxy properties, we first determine the direction along which each \galaxpy parameter modifies colours. Then, we proceed to determine the angle between each of these directions and the SFH-direction. A small (large) angle between the SFH-direction and a certain galaxy property indicates that variations in both this property and the SFH have a similar (different) impact on galaxy colors. To determine the direction associated with a particular \galaxpy parameter in colour space, we start by computing, for each SFH from any of the libraries introduced in \S\ref{sub:data_sfh_library}, the first eigenvector of the distribution of colours resulting from varying the target parameter while holding all other parameters fixed to their fiducial values. We thus end up with a distribution of eigenvectors, one for each element of the \umachine and \illustris libraries. Then, we marginalise over SFH variations by taking the average of this collection of eigenvectors; we refer to the resulting vector as the direction associated with the \galaxpy parameter under consideration. It is worth noticing that the first PC of all galaxy properties but $Z$ explains more than 97\% of the variance; for $Z$, the variance explained by the 1st PC drops to 90\% (see Table~\ref{tab:extra_params}).

In Fig.~\ref{fig:diff_params}, orange arrows denote the direction along which distinct galaxy properties modify colours, while blue arrows show the SFH-direction. In Table~\ref{tab:extra_params} we collate the angle between the SFH-direction and the direction associated with each \galaxpy parameter; we find that all these angles are larger or equal to 12 degrees. {\em We thus conclude that changes in metallicity, dust attenuation, and nebular emission line ratios modify colours in distinct directions of the colour space relative to SFH variations}.

\subsection{Impact of galaxy properties on the SFH-direction}
\label{subsec:sfh_direction_robustness}

Before presenting the influence of each individual galaxy feature on broad-band colours, we first study whether the orientation of the SFH-direction depends on such properties. For the sake of definiteness, we begin by using the IMF as an example. We first select the colours generated for each SFH in our libraries using different values of the target parameter while holding the other \galaxpy parameters fixed to their fiducial values. We group these colours according to the value considered for the target parameter, in this particular case a Chabrier, Salpeter, and Kroupa IMF, ending up with three distinct sets of colours. We then compute the first eigenvector of each of these colour distributions. Note that, by definition, the first eigenvector associated with Chabrier colours is the SFH-direction. Finally, we compute the angles between the Salpeter and Kroupa eigenvectors and the SFH-direction. In Table~\ref{tab:extra_params} we show the median of these two angles, which we use to give an estimate for how much the orientation of the SFH-direction changes due to physically plausible variations in the IMF. We repeat the same process for each \galaxpy parameter, gathering the results in Table~\ref{tab:extra_params}. In all cases, we find that the SFH-direction never varies by more than 9 degrees owing to variations in other galaxy features. Based on this exercise, we conclude that the orientation of the SFH-direction is very robust: {\em no matter the properties of the galaxy, the formation of stars smoothly varies broad-band colours along the same direction in colour space.}


\subsection{Initial Mass Function}

The initial mass function is a significant source of systematic uncertainty in galaxy evolution \citep[for a recent review, see][]{Hopkins2018}. The impact of IMF variations on the broad-band colours of single stellar populations is rather small, as SSP colours are dominated by stars with masses close to the main sequence turnoff \citep[Figure 7,][]{Conroy2009}. However, it is conceivable that the IMF may modify the colours of {\em composite} stellar populations due to the large range of turnoff masses contributing to the integrated light. To study the influence of the IMF upon broad-band colours, we leverage physically-motivated SFHs from \umachine and \illustris, and we use \galaxpy to generate colours assuming Chabrier, Kroupa, and Salpeter IMFs. We find that broad-band colours have very little dependence on the IMF for realistic composite stellar populations; the median difference between u-g and z-Y colours produced using a Chabrier and a Kroupa (Salpeter) IMF is 2.4 and 0.3 mmag (18.4 and 5.7 mmag), respectively. Consequently, the impact of IMF variations on broad-band colours is very weak. Nonetheless, these small differences shift colours in the same direction for all galaxies. Intuitively, Salpeter and Kroupa IMFs result in redder colours relative to a Chabrier IMF because the first two are characterised by a larger ratio of low-to-high mass stars than the latter.


\subsection{Metallicity}

It is well-known that the temperature and luminosity of main sequence stars are inversely proportional to their metallicity \citep[e.g.,][]{Bruzual2003}; galaxies therefore get redder as their average metallicity increases \citep[e.g.,][for results from observations]{Sanders2013}. By analysing the colours generated using SFHs from \umachine and \illustris, we confirm that sub-solar (super-solar) metallicities result in broad-band colours that are bluer (redder) than those corresponding to solar metallicities. Regarding the size of these variations, the median difference between u-g and z-Y colours produced using a solar and a sub-solar $Z=0.004$ metallicity (super-solar $Z=0.05$ metallicity) is -0.33 and -0.16 mag (0.07 and 0.09 mag), respectively, where positive and negative values indicate a shift to the red and blue. Furthermore, the direction of colour fluctuations owing to metallicity differences, i.e., the metallicity-direction, presents an angle of 35 degrees with respect to the SFH-direction. Taken together, these results show that metallicity variations have a strong effect on galaxy colours, especially on those with shorter wavelengths, and that changes in metallicity modify colours in a completely distinct direction in colour space relative to SFH variations.


\subsection{Dust attenuation}

Dust grains absorb UV and optical light, and then they re-radiate this energy in the infrared; consequently, galaxies get redder as their dust content increases. It is worth noticing that even though the attenuation of starlight is indeed directly proportional to the total amount of dust in a galaxy, the precise shape of the attenuation curve depends upon dust properties such as the distribution of grain sizes \citep[e.g.,][]{Weingartner2001}. In \galaxpy we use the \citet{Charlot2000} dust model, which includes one parameter controlling the attenuation of light from coming from stars younger than 10 Myr, $\tau_{V,{\rm y}}$, and one for stars older than this age, $\tau_{V,{\rm o}}$. Broadly, we find that the impact of each parameter is similar, making broad-band colours redder. The median difference between u-g and z-Y colours produced using $\tau_{V,{\rm y}}=0$ and 3 ($\tau_{V,{\rm o}}=0$ and 1) is 0.32 and 0.06 mag (0.19 and 0.06 mag), respectively. Dust attenuation has thus a strong impact on broad-band colours, especially on those with shorter wavelengths.

As with the other \galaxpy parameters, we compute the direction in colour space associated with variations in $\tau_{V,{\rm y}}$ and $\tau_{V,{\rm o}}$. We find that the $\tau_{V,{\rm y}}$- and $\tau_{V,{\rm o}}$-directions present an angle of 12 and 27 degrees with respect to the SFH-direction, respectively. Therefore, the SFH- and $\tau_{V,{\rm y}}$-direction are rather similar, a manifestation of the well-known age-dust degeneracy \citep[e.g.,][]{Papovich2001}. On the other hand, the angle formed by the $\tau_{V,{\rm o}}$- and SFH-directions is significantly larger, and so evidently the attenuation of light from old stars leaves a signature on broad-band colours that is discernible from the influence of star formation history. 

Finally, we study the dependence of the direction associated with dust attenuation on the shape of the attenuation curve using the Calzetti \citep{calzetti00} and Fitzpatrick \citep{Fitzpatrick1999} laws. In particular, we generate colours for each of these laws using ten linearly-spaced values for the total V-band attenuation spanning the interval $A_V \in [0,3]$ mag. We find that the first eigenvector of each of these two attenuation laws captures more than 99.99\% of the variance in the data and that the Calzetti (Fitzpatrick) eigenvector presents an angle of 20 (21) degrees with the SFH-direction. These results let us conclude that the impact of the shape of the attenuation curve on the direction along with dust attenuation modify colours is weak. In fact, the Calzetti and Fitzpatrick eigenvectors present an angle of just 8 and 6 degrees with the $\tau_{V,{\rm o}}$-direction, respectively.


\subsection{Nebular emission}

The impact of nebular emission lines on broad-band colours is more important for galaxies with low metallicities, large SFRs, and at high redshift \citep[e.g.,][]{Conroy2013}. The influence of nebular emission on any particular colour has a dramatic dependence on redshift because the filter in which a specific emission line falls changes as the central wavelength of such line increases with redshift. The introduction of nebular emission lines thus results in some broad-band colours getting redder while others bluer; this is in contrast to variations in IMF, metallicity, and dust attenuation, because changes in these other properties systematically shift all colours towards the blue or red. In what follows, we first study the impact of nebular emission lines on broad-band colours at $z=0.03$; we then address the effect of considering different emission line ratios.

In \galaxpy, the luminosity of nebular emission lines is proportional to the star formation rate of a galaxy; therefore, we expect the colours of star-forming galaxies to be more strongly influenced by nebular emission than those of quenched galaxies. Indeed, the median difference between u-g and g-r colours produced with and without considering emission lines for star-forming (quenched) galaxies is -44.7 and 22.1 mmag (-9.7 and 1.1 mmag), respectively, where we model emission lines using $Z_{\rm ISM}=0.001$ and $\log_{10}U_S=-3.5$. This general trend can be seen in the bottom-middle and bottom-right panels of Fig.~\ref{fig:diff_params}; galaxies with bluer (redder) colours are more (less) affected by nebular emission lines. On the other hand, the impact of nebular emission lines on broad-band colours is weaker relative to that of both metallicity and dust attenuation.

There are two parameters controlling nebular emission line ratios in \galaxpy, $Z_{\rm ISM}$ and $U_S$. In contrast to other \galaxpy parameters, a monotonic change any of these two parameters does not result in broad-band colours getting systematically bluer or redder. Consider, for example, how the colours of a galaxy change when excluding emission lines altogether versus including emission lines using $Z_{\rm ISM}=0.001$ and $\log_{10}U_S=-1$ or the same value of $Z_{\rm ISM}$ and $\log_{10}U_S=-4$. We find that for the first case the u-g and g-r colours shift by 19.4 and -13.2 mmag, respectively, while for the second case the variations are -35.6 and 15.1 mmag.

Some combinations of $Z_{\rm ISM}$ and $U_S$ produce emission line ratios that are incompatible with observations, and thus the colour distributions shown in the bottom-middle and bottom-right panels of Fig.~\ref{fig:diff_params} overestimate the range of colour variations expected for realistic nebular emission lines. Regarding the directions in colour space associated with variations in these two parameters, we find that the angles formed by the $Z_{\rm ISM}$- and $U_S$-directions and the SFH-direction are 22 and 55 degrees, respectively. Nebular emission lines thus vary galaxy colours in quite distinct directions in colour space relative to SFH variations. Finally, we note that nebular continuum emission is currently neglected in \galaxpy; as a result, star-forming galaxies should present even bluer colours and the directions associated with nebular emission line parameters could be rotated.


\begin{figure}
	\includegraphics[width=\columnwidth]{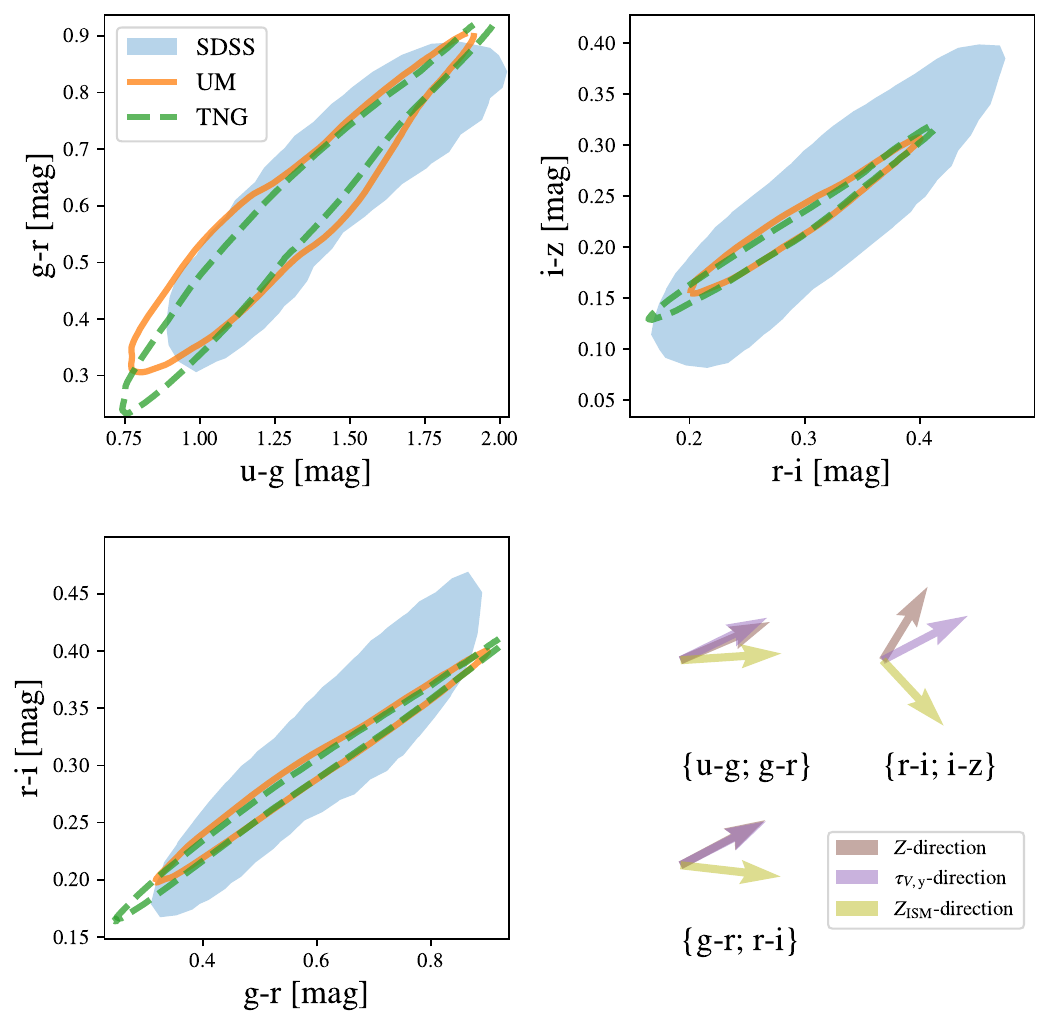}
    \caption{\label{fig:SDSS_cols} Colours of SDSS galaxies and theoretical predictions for the distribution of colours that results from physically-motivated SFH variations. Shaded areas display the results for SDSS galaxies, while solid (dashed) contours show \umachine (\illustris) predictions. Both contours and shaded areas enclose 90\% of the colour distributions. Brown, purple, and yellow arrows indicate the direction in colour space associated with changes in metallicity, dust attenuation, and nebular emission lines, respectively. The major axis of the SDSS ellipsoid is closely aligned with the major axes of the \umachine and \illustris ellipsoids, while it presents a significant angle with respect to the directions associated with variations in other galaxy properties. Taken together, these results suggest that the SFH is the most important driver of optical and near-infrared galaxy colours.}
\end{figure}

\section{Analysis of broad-band colours from observations}
\label{sec:colorsobv}

In \S\ref{sec:SFHSED}, we used PCA to study the influence of SFH on broad-band colours, finding that physically-motivated SFH variations modify colours along a single direction in colour space, the SFH-direction. Then, in \S\ref{sec:colorsgprop}, we conducted a similar analysis of other galaxy features, finding that changes in the IMF, metallicity, dust attenuation, and nebular emission move galaxy colours along distinct directions in colour space relative to the SFH-direction. In this section, we seek to identify the galaxy feature(s) with the strongest influence on the colours of observed galaxies. To do so, we again employ the PCA-based methodology used in \S\ref{sec:SFHSED} and \ref{sec:colorsgprop}, only here we analyse the broad-band colours of a volume-limited sample of galaxies from the Sloan Digital Sky Survey \citep[SDSS,][]{york00, dawson13}. Throughout this section we only consider the four colours resulting from subtracting consecutive SDSS bands: \{u-g, g-r, r-i, i-z\}.

We proceed to analyse the broad-band colours of galaxies from the SDSS data release 10 \citep{Ahn2014} with secure spectroscopic redshifts within the narrow interval $0.025<z<0.035$ and with stellar masses greater than $\log_{10}(\Mstar[\Msun])=9.5$; the first and second criterion are introduced to minimise the impact of redshift evolution and to avoid completeness issues, respectively. We start by carrying out a PCA on the colours of the galaxies in this sample, finding that the first and second PCs explain 89.2 and 8.7\% of the variance, respectively. The colours of SDSS galaxies thus span a multi-dimensional ellipsoid with significant extent along two independent dimensions. 

In Fig.~\ref{fig:SDSS_cols} we compare the distribution of broad-band colours of SDSS galaxies alongside theoretical expectations for the distributions that result from variations in SFH alone. Blue shaded areas display the distribution of SDSS colours, while orange solid (green dashed) contours present the distribution of simulated colours produced by variations in SFH predicted by \umachine (\illustris) (see \S\ref{sec:SFHSED}). In particular, the volume spanned by the orange and green contours is exclusively due to variations in SFH, as we hold all other galaxy properties fixed to their fiducial values. Both shaded areas and contours enclose 90\% of the colour distributions. 

As expected, the major axes of the \umachine and \illustris ellipsoids are closely aligned, supporting the thesis that the SFH-direction is universal. Additionally, the breadth of the orange and green ellipsoids is very similar, confirming that physically-motivated SFH variations result in galaxy colours spanning a similar volume in colour space. Furthermore, the major axes of simulated and observed colours also appear to be well-aligned, suggesting that the most important galaxy property controlling broad-band colours is the star formation history. More quantitatively, we find that the angle between the SFH-direction and the first eigenvector of SDSS colours is very small, approximately 7 degrees, comparable to the small size of changes to the SFH-direction induced by variations in distinct galaxy properties (see \S\ref{subsec:sfh_direction_robustness}). On the other hand, the angles between the first eigenvector of SDSS galaxies and the IMF-, $Z$-, $\tau_{\rm V, y}$-, $\tau_{\rm V, o}$-, $Z_{\rm ISM}$-, and $U_S$-directions are 20, 27, 15, 31, 15, and 47 degrees, respectively. Taken together, these results suggest that {\em broad-band optical and near-infrared galaxy colours are primarily driven by star formation history}.

Despite the close correspondence of the SFH-direction with the first eigenvector of SDSS colours, in Fig.~\ref{fig:SDSS_cols} we can also see that the ellipsoid of SDSS colours spans a much larger volume than that of \umachine and \illustris. Therefore, it is clear that SFH variations alone cannot completely account for the colours of observed galaxies. To help visualise this, in the bottom right panel of this figure we display the directions associated with variations in metallicity, dust attenuation, and nebular emission lines. We can see that these are indeed the directions needed to stretch the ellipsoids of simulated colours in order to cover the volume of colour space spanned by observed galaxies. Thus, even though SFH is the primary driver of galaxy colours, variations in metallicity, dust attenuation, and nebular emission lines are nonetheless required in order to account for the colours of observed galaxies.


\section{Discussion}
\label{sec:discussion}

Determining the physical characteristics of individual galaxies is one of the most common applications of SPS modelling, and there is a diversity of methodologies that have been deployed to infer SFHs in particular. Broadly speaking, all such analyses proceed by programmatically varying a SFH model together with other ingredients such as metallicity and dust attenuation. The observational predictions of each model are then compared to corresponding measurements of the galaxies of interest, enabling determination of the best-fit point (and sometimes confidence intervals) on the model parameter space. This approach to SFH inference has been explored using a wide variety of observations, including individual spectral features \citep{kauffmann_etal03, Chauke2018}, broad-band photometry \citep{Conroy2009}, and colour-magnitude diagrams of resolved stellar populations \citep{McQuinn2010} and unresolved stellar populations \citep{cook_etal19}. See \citet{Conroy2013} for a comprehensive review. 

A natural way to classify the landscape of SFH models is to divide them into parametric and non-parametric approaches. In parametric models, SFHs are described by a family of functions controlled by a set of parameters that are explored during the analysis. This approach is computationally efficient and transparent, and it has been shown that even relatively simple models can capture most SFHs predicted by simulations with reasonably high fidelity \citep{simha_etal14, diemer_etal17}. Common choices for parameterized SFHs include single component $\tau$-models \citep{Schmidt1959}, rising SFHs aiming to explain the SED of high-redshift galaxies \citep{Buat2008, Lee2009}, log-normal SFHs motivated by measurements of cosmic star formation rate density \citep{Gladders2013}, and various two-epoch models with separate rising and declining phases at early and late times \citep{Ciesla2017, tinker17, Carnall2018}. It is important to note that none of these functional forms can account for complex features such as episodic bursts of star formation, or rejuvenation following a sudden quenching event. Therefore, the simplicity of these models may bias SFH inference for galaxies with such features.

Alternatively, non-parametric models do not explicitly specify an analytical function for star formation history. Instead, they typically define SFHs based on SFRs computed at a finite set of control points. A wide range of such SFH models have been explored in the literature, including those based on piecewise constant functions \citep{CidFernandes2005, Ocvirk2006} and models with time-stepping that is adaptive \citep{tojeiro_etal07, iyer_etal19} and/or stochastically correlated \citep{caplar_tachella19}. As expected, nonparametric models can predict a broader variety of features in the SFH relative to a family of simple analytical functions. However, this added flexibility usually comes at significant computational cost, and potentially encompasses freedom to capture features in SFH that are either not physical, not warranted by the available data, or both. After all, there exists a finite amount of SFH information contained in any observational dataset, and so one must be careful not to overfit limited data with a highly complex model. Even with high-resolution spectra in optical wavelengths, at most $\sim8$ episodes of star formation can be reliably recovered \citep{Ocvirk2006}. Despite photometric measurements through a few broad bands contain considerable constraining power on galaxy parameters \citep{Pacifici2016}, much information about the SFH is lost when observing a galaxy's SED through only a handful of broad-band filters. As a result, imaging data alone constrain even fewer SFH features relative to spectroscopic data \citep{Carnall2019}.

To address the ill-conditioned nature of SFH inference with a flexible non-parametric model, it is typical to appeal to physically motivated priors to reduce the dimensionality of the problem. In one approach to applying such priors, galaxy catalogues generated by semi-analytic models or hydrodynamical simulations are treated as libraries from which SFHs are drawn \citep{Finlator2007, Brammer2008, Pacifici2012}. Synthetic galaxy catalogues can also be used as training data for a machine learning algorithm such as a Convolutional Neural Network \citep{lovell_etal19}, or for a flexible basis function decomposition \citep{iyer_gawiser17, iyer_etal19}.\footnote{Such catalogs are also widely used to {\em validate}, rather than train, techniques for inferring physical galaxy properties, e.g. \citet{Laigle2019}.}

Using a library of SFHs from the semi-analytic model developed in \citet{delucia_blaizot07}, \citet{Pacifici2015} studied how changes in numerous modelling ingredients propagate into variations in colours. \citet{Pacifici2015} showed that even the full panoply of SFHs predicted by the SAM is insufficient to reproduce the diversity of observed galaxy colours. One of the most important conclusions of the present work is that physically-motivated SFHs result in galaxy colours tightly localised along a single direction in colour space. Our identification of this ``SFH-direction" reinforces the results presented in \citet{Pacifici2015}, in addition to providing confirmation with SFHs drawn from two independent models. Moreover, we show that the SFH-direction is closely aligned with the first eigenvector of SDSS galaxies, suggesting that galaxy colours are mainly driven by star formation history.


\section{Conclusions}
\label{sec:conclusions}

The spectral energy distribution of a galaxy results from the complex interaction of many factors, including its star formation history, metallicity evolution, and dust properties. In this work, we take advantage of the newly developed galaxy spectral prediction tool \galaxpy to study the influence of distinct galaxy properties on broad-band optical and near-infrared colours from 3\,200 to 10\,800 {\AA} at $z\simeq0$. Our main findings can be summarised as follows:

\begin{itemize}

\item Using SFHs predicted by the empirical model \umachine and the cosmological hydrodynamical simulation \illustris, we show that physically motivated SFH variations modify broad-band colours along a single direction in colour space, the SFH-direction. We find that the SFH-direction is universal, i.e. it is the same for galaxies in haloes of different masses, independently of whether these are star-forming or quenched galaxies, as well as for the two models studied. Furthermore, we show that variations in any other galaxy property have very little impact on the SFH-direction.

\item We determine that the projection of optical and near-infrared galaxy colours onto the SFH-direction is mostly regulated by the fraction of stellar mass formed over the last billion years. Taken together in combination with the phenomenon of downsizing, i.e., the fact that galaxies in more massive haloes reach the peak of their SFH faster, have greater stellar mass, and quench earlier, this result accounts for the observation that galaxies get increasingly redder as their host haloes become more massive.

\item We show that the broad-band colours of low redshift SDSS galaxies span a multidimensional ellipsoid with significant extent along two independent dimensions. The major axis of this ellipsoid is closely aligned with the SFH-direction, while it presents a significant angle with respect to the directions associated with variations in other galaxy properties. Considering together, these results suggest that the main driver of optical and near-infrared galaxy colours is the star formation history. Furthermore, we show that despite broad-band colours are mostly controlled by SFH, changes in metallicity, dust attenuation, and nebular emission lines are also required to explain the diversity of galaxy colours from observations.

\end{itemize}

In a companion paper to the present work, we will exploit the simplicity of the influence of SFH upon galaxy colours to create a surrogate model for broad-band photometry. By leveraging high-performance implementations of machine-learning algorithms on GPU resources, the resulting emulator will create the capability to map approximate, but high-accuracy colours to survey-scale volumes of simulated galaxies at a tiny fraction of the time taken by traditional SPS codes. These two works will thus serve as the foundation of Surrogate modelling the Baryonic Universe (SBU), a new simulation-based method to forward-model the galaxy-halo connection.


\section*{Acknowledgements}

We acknowledge the anonymous referee for helpful comments that have made the paper both clearer and stronger. We thank Matt Becker, Peter Behroozi, Andrew Benson, Benedikt Diemer, Eve Kovacs, and Alba Vidal-Garc\'ia for useful discussions. Argonne National Laboratory's work was supported by the U.S. Department of Energy, Office of Science, Office of Nuclear Physics, under contract DE-AC02-06CH11357. We gratefully acknowledge use of the Bebop cluster in the Laboratory Computing Resource Center at Argonne National Laboratory. Computational work for this paper was also performed on the Phoenix cluster at Argonne National Laboratory, jointly maintained by the Cosmological Physics and Advanced Computing (CPAC) group and by the Computing, Environment, and Life Sciences (CELS) Directorate.

Funding for SDSS-III has been provided by the Alfred P. Sloan Foundation, the Participating Institutions, the National Science Foundation, and the U.S. Department of Energy Office of Science. The SDSS-III web site is \url{http://www.sdss3.org/}. SDSS-III is managed by the Astrophysical Research Consortium for the Participating Institutions of the SDSS-III Collaboration including the University of Arizona, the Brazilian Participation Group, Brookhaven National Laboratory, Carnegie Mellon University, University of Florida, the French Participation Group, the German Participation Group, Harvard University, the Instituto de Astrof\'isica de Canarias, the Michigan State/Notre Dame/JINA Participation Group, Johns Hopkins University, Lawrence Berkeley National Laboratory, Max Planck Institute for Astrophysics, Max Planck Institute for Extraterrestrial Physics, New Mexico State University, New York University, Ohio State University, Pennsylvania State University, University of Portsmouth, Princeton University, the Spanish Participation Group, University of Tokyo, University of Utah, Vanderbilt University, University of Virginia, University of Washington, and Yale University.


\bibliographystyle{mnras}
\bibliography{biblio}

\appendix
\renewcommand{\thefigure}{A\arabic{figure}}

\begin{figure}
    \begin{center}
	\includegraphics[width=0.85\columnwidth]{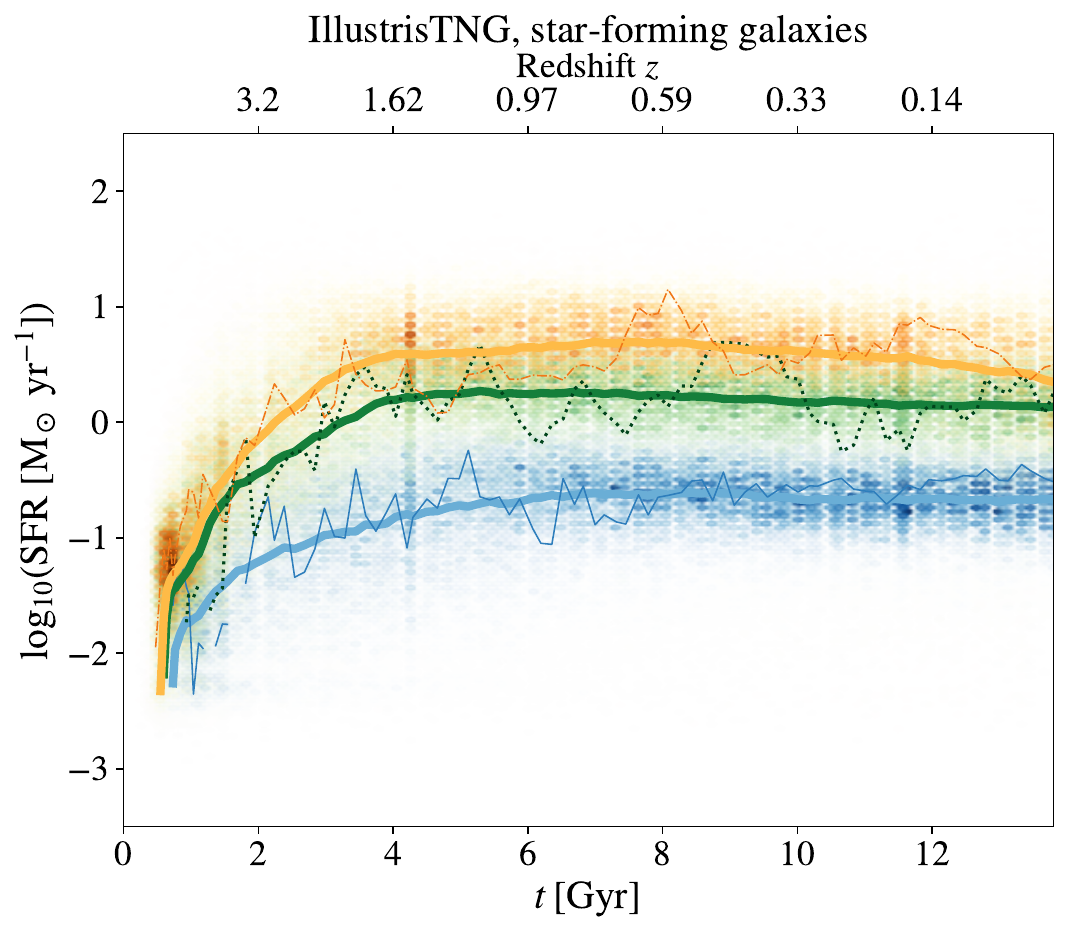}

	\includegraphics[width=0.85\columnwidth]{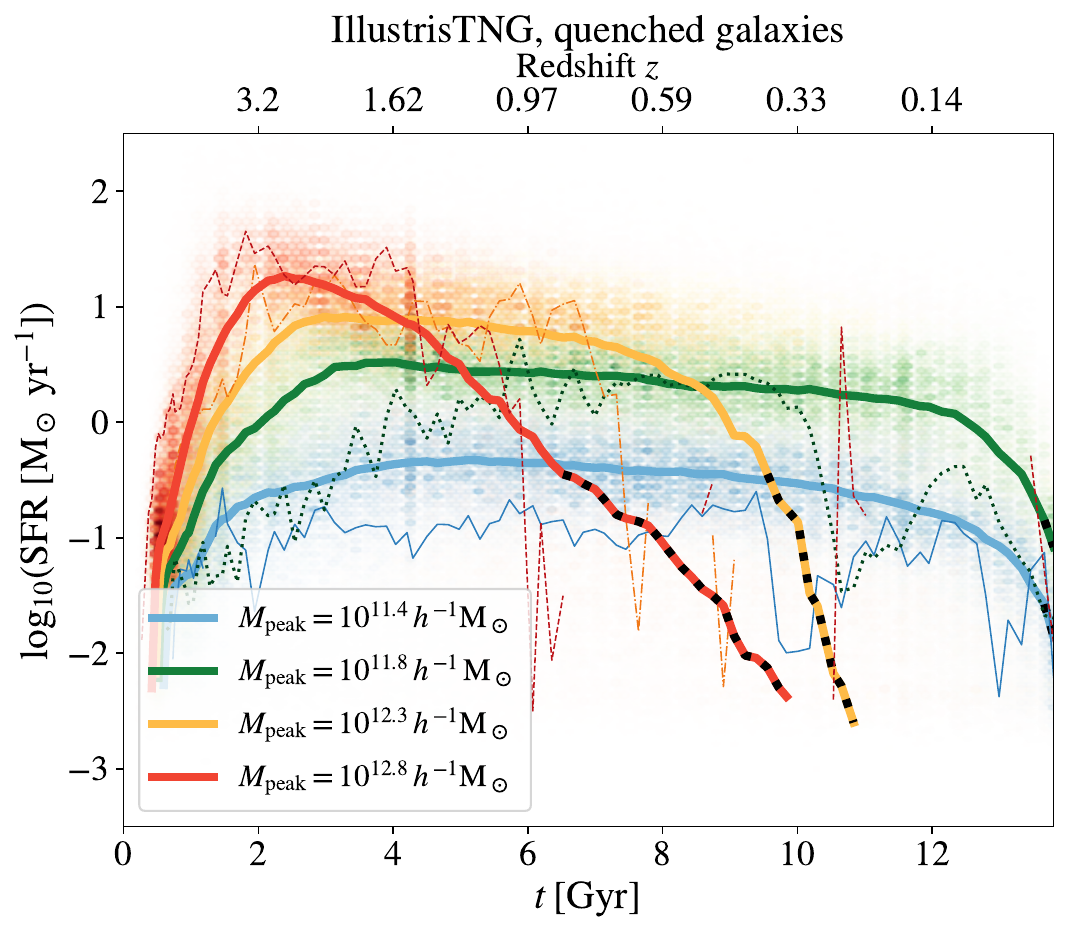}
    \end{center}

    \caption{\label{fig:SFH_TNG} Same as Fig.~\ref{fig:SFH_Umachine} for SFHs drawn from \illustris. The main difference between SFHs predicted by \umachine and \illustris is that after the peak of SFR, the latter decrease more slowly (rapidly) for star-forming (quenched) galaxies.}
\end{figure}

\begin{figure}
    \begin{center}
        \includegraphics[width=\columnwidth]{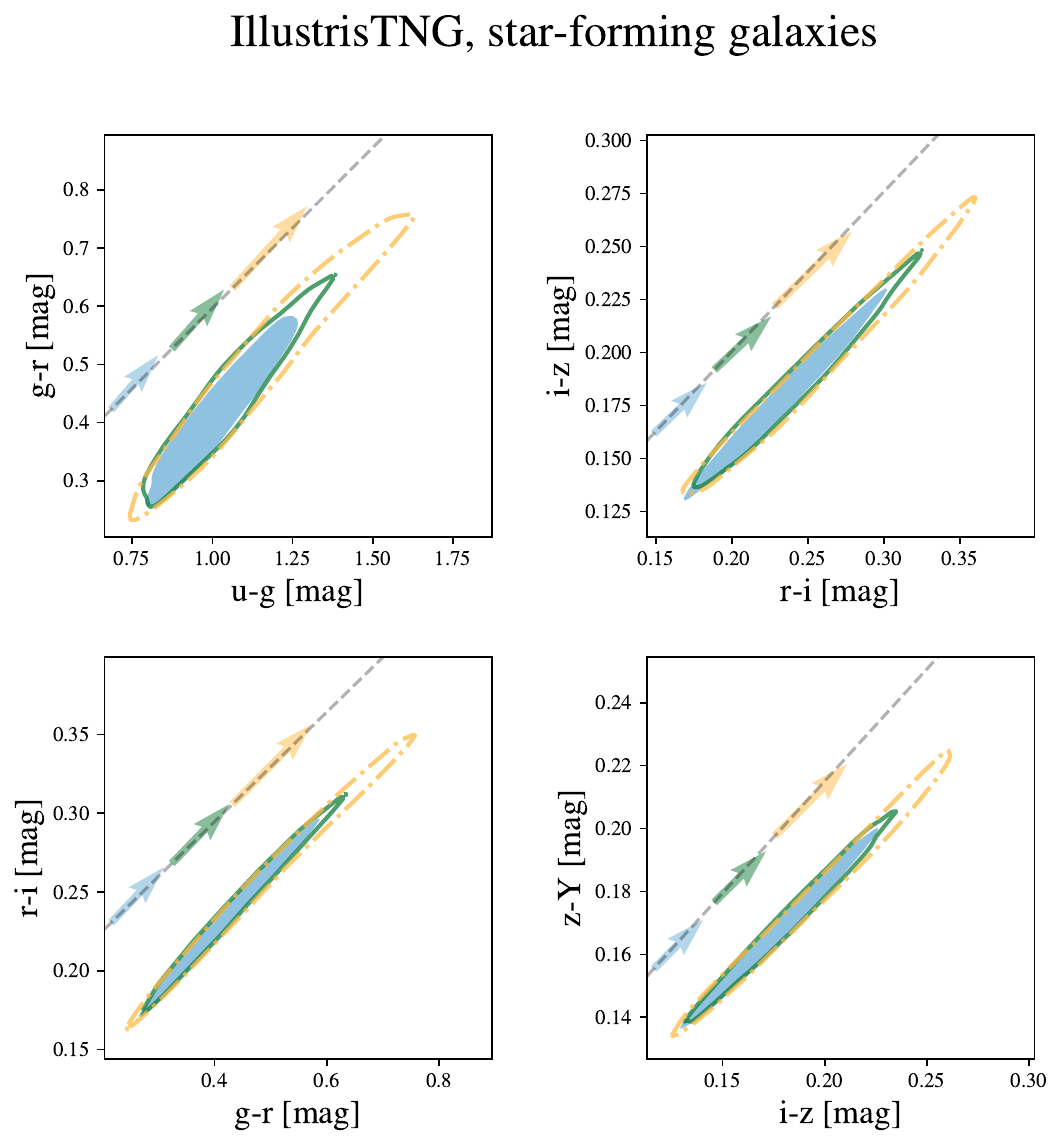}
        
        \includegraphics[width=\columnwidth]{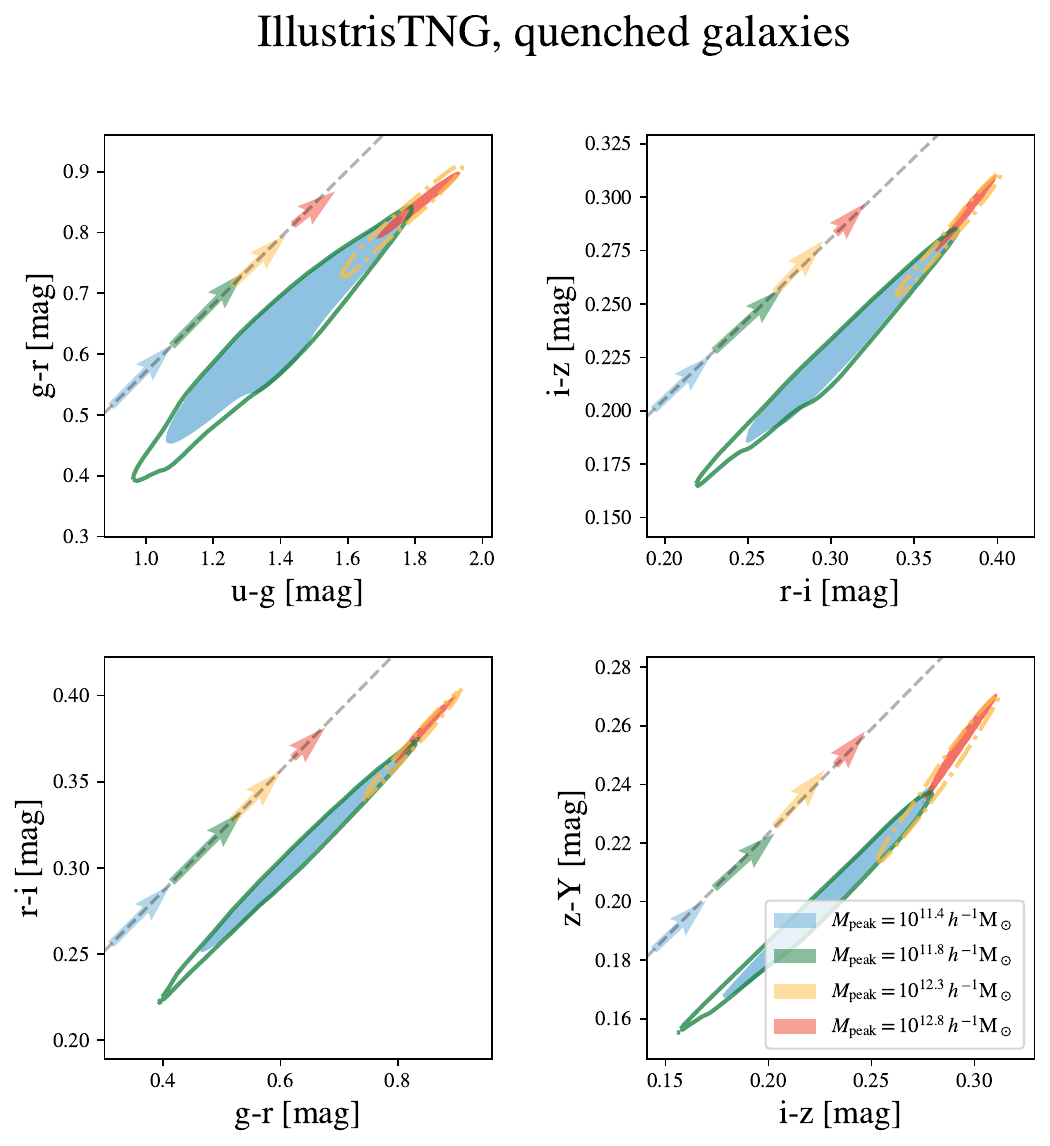}
    \end{center}
    \caption{\label{fig:SFH_colors_TNG} Same as Fig.~\ref{fig:SFH_colors} for colours produced using \illustris SFHs. Even though SFHs predicted by \umachine and \illustris have different shapes and levels of burstiness, the resulting colour distributions are very similar.}
\end{figure}

\begin{figure}
	\includegraphics[width=\columnwidth]{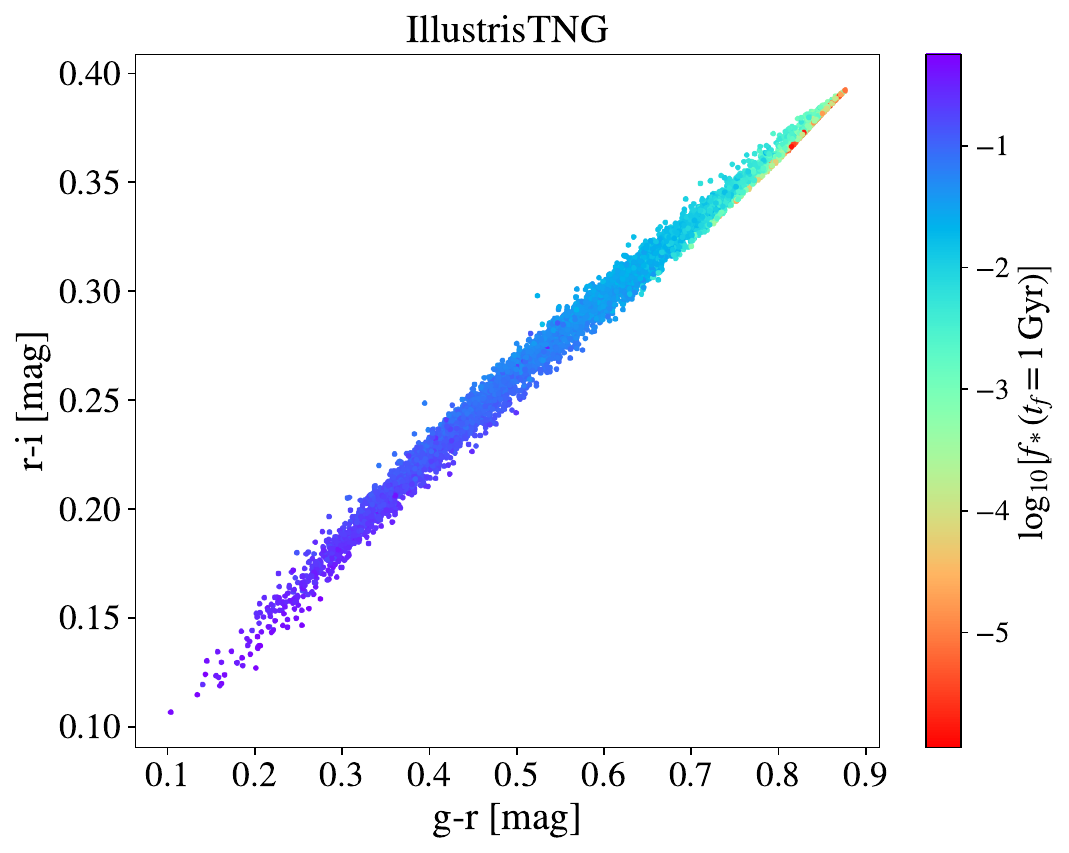}
    \caption{\label{fig:colordif_IM_TNG} Same as Fig.~\ref{fig:colordif_IM} for colours produced using SFHs predicted by \illustris. We can see that \fstar is not only strongly correlated with \umachine colours, but also with \illustris colours.}
\end{figure}

\section{Additional figures for IllustrisTNG}
\label{app:tng_res}

For brevity, throughout the main body of this work we mostly included figures presenting results for \umachine. Here, we display analogous figures for \illustris.

In Fig.~\ref{fig:SFH_TNG} we show SFHs predicted by \illustris for star-forming and quenched galaxies (see Fig.~\ref{fig:SFH_Umachine} for \umachine). We can clearly see that the SFHs predicted by these two models present different shapes. First, after cosmic star formation peaks at $z\sim 1-3$, the SFR of star-forming galaxies in \umachine decreases more rapidly relative to \illustris. Second, at fixed halo mass, galaxies in \illustris become quenched at earlier times relative to \umachine. Finally, we also find that SFHs drawn from \umachine exhibit substantially more burstiness relative to those from \illustris.

Although SFHs predicted by \umachine and \illustris have different shapes and levels of burstiness, in both models the influence of SFH variations on colours is very similar. In Fig.~\ref{fig:SFH_colors_TNG} we display the distribution of galaxy colours resulting from the diversity of SFHs predicted by \illustris (see Fig.~\ref{fig:SFH_colors} for \umachine). As in \umachine, SFH variations result in galaxy colours moving along a single direction in colour space. Because the modelling of SFH in these two models is predicated upon quite distinct assumptions, there is good reason for confidence that the universality of the SFH-direction is robust to systematic uncertainties in the true physical processes that regulate star formation.

In Fig.~\ref{fig:colordif_IM_TNG} we show the correlation between broad-band colours and \fstar at $z=0.03$ in \illustris (see Fig.~\ref{fig:colordif_IM} for \umachine). We find that the projection of galaxy colours onto the SFH-direction is strongly correlated with \fstar in both \umachine and \illustris. Again, given the differences between SFHs predicted by \umachine and \illustris, it is remarkable that \fstar captures so precisely the influence of SFH variations on galaxy colours.

\bsp
\label{lastpage}
\end{document}